\documentclass[amsmath,amssymb,aps,pra,twocolumn,notitlepage,superscriptaddress,longbibliography]{revtex4-2}
\usepackage{graphicx}
\usepackage{amsfonts}
\usepackage{amssymb}
\usepackage{amsthm}
\usepackage{array}
\usepackage{amsmath}
\usepackage{verbatim} 
\usepackage{hyperref}   
\usepackage{xcolor}
\usepackage{bbold}
\usepackage{epstopdf}
\usepackage{mathtools}
\usepackage[normalem]{ulem}
\usepackage{physics}
\usepackage{soul}

\newtheorem{theorem}{Theorem}

\newtheorem{corollary}{Corollary}[theorem]

\def\bra#1{\left< #1\right|}
\def\ket#1{\left|#1 \right>}

\def\Tr{\mbox{Tr}}

\newcommand{\del}[1]{{\iffalse #1 \fi}}

\def\draft{1}
\ifnum\draft=1
\newcommand{\han}[1]{\textcolor{blue}{ Minki: #1}}
\else 
\newcommand{\han}[1]{}
\fi

\begin{document}
\title{Trade-off between Information Gain and Disturbance in Local Discrimination of Entangled Quantum States}

\author{Youngrong Lim}
\affiliation{School of Computational Sciences, Korea Institute for Advanced Study, Seoul 02455, Korea}
\affiliation{Department of Physics, Chungbuk National University, Cheongju, Chungbuk, 28644, Korea}
\author{Minki Hhan}
\affiliation{Quantum Universe Center, Korea Institute for Advanced Study, Seoul 02455, Korea}
\author{Hyukjoon Kwon}
\email{hjkwon@kias.re.kr}
\affiliation{School of Computational Sciences, Korea Institute for Advanced Study, Seoul 02455, Korea}

\begin{abstract}
We establish an information gain-disturbance trade-off relation in local state discrimination. Our result demonstrates a fundamental limitation of local strategy to discriminate entangled quantum states without disturbance, which becomes more difficult as the entanglement of the states to be discriminated increases. For a set of maximally entangled states, the capability of local strategy is tightly suppressed, as random guessing without measurements saturates the bound provided by the trade-off relation. We also show that the trade-off can be circumvented when local operations are aided by pre-shared entanglement. To simultaneously achieve correct guessing of state and non-disturbance, an entirely different strategy from conventional state discrimination should be adopted to lower the cost of pre-shared entanglement. We explicitly propose an adaptive and non-destructive strategy based on the stabilizer formalism, which shows a strict advantage over conventional teleportation-based approaches in pre-shared entanglement cost for discriminating a set of maximally entangled states. As an application of the trade-off relation, we propose an entanglement certification protocol that is robust against depolarizing noise and generalize it to multipartite scenarios in a quantum network.

\end{abstract}
\maketitle

\section{Introduction}
Quantum state discrimination~\cite{helstrom1969quantum, holevo1974remarks, barnett2009quantum, bae2015quantum} has served as a primitive for exploring the foundations of quantum theory, including the no-cloning theorem~\cite{wootters1982single}, nonlocality~\cite{bennett1999quantum}, and contextuality~\cite{schmid2018contextual}. 
In particular, when two physically separated parties are only allowed to perform local operations and classical communications (LOCC) to discriminate quantum states, often referred to as \textit{local state discrimination} (LD), even orthogonal states cannot be definitively distinguished~\cite{bennett1999quantum, ghosh2001distinguishability}. Intriguing observations via LD have deepened our understanding of fundamental concepts in quantum theory, such as quantum nonlocality without entanglement~\cite{bennett1999quantum} and operational resource theory of imaginarity~\cite{wu2021operational}.
Based on these observations, various applications in data hiding~\cite{terhal2001hiding, divincenzo2002quantum, eggeling2002hiding, matthews2009distinguishability}, secret sharing~\cite{markham2008graph,rahaman2015quantum,banik2021multicopy}, and quantum error correction~\cite{kribs2019quantum} have been developed.

Nevertheless, characterizing the limitation of LD for a given set of states remains a non-trivial problem. Since there exists a set of product states that cannot be distinguished using LOCC, known as quantum nonlocality without entanglement~\cite{bennett1999quantum}, the entanglement of the states to be distinguished is not directly connected to the limitation of LD. Moreover, any two maximally entangled states (MESs) are perfectly distinguishable by LOCC, whenever they are orthogonal to each other~\cite{walgate2000local}. Despite these examples, extensive investigations have consistently demonstrated a prevailing trend that the ability of LOCC is highly restricted when discriminating entangled quantum states~\cite{ nathanson2005distinguishing,yu2012four, nathanson2013three,10.5555/2685164.2685167,li2015d, tian2016classification,lugli2020fermionic,yuan2022finding}. However, these results heavily rely on the choice of a set of states, leaving the complete understanding of their relationship with entanglement still open.

Meanwhile, quantum state discrimination necessarily involves measurements on the states to gain which-state information. The basic principle of quantum mechanics tells us that a quantum state can be disturbed by the impact of measurement. Hence, the state after the discrimination process can be different from the original input. Especially when it comes to local discrimination of entangled quantum states, disrupted entanglement after the discrimination process cannot be recovered by LOCC~\cite{bennett1996concentrating}. Given the practical relevance of conserving and reusing entanglement in subsequent information processing~\cite{kim2012protecting,doi:10.1126/sciadv.adi5261}, it is important to understand an explicit trade-off relation between information gain and state disturbance. Although such trade-offs have been extensively studied from the fundamental and practical points of view~\cite{fuchs1996quantum,banaszek2001fidelity,buscemi2006information,sacchi2006information,buscemi2008global,kretschmann2008information,hong2022demonstration}, they have not been fully investigated in the context of state discrimination, except for specific cases: state discrimination without local strategies~\cite{buscemi2006information,skrzypczyk2019robustness} and LD with a strict constraint, such as entanglement preservation~\cite{cohen2007local}.

In this manuscript, we establish a trade-off between distinguishability and disturbance in LD of entangled states. We introduce a score function that simultaneously captures the correct guessing probability and the state disturbance after the discrimination process. An upper bound on the score function is given in terms of the entanglement of states to be discriminated. This implies that more entanglement renders a tighter trade-off relation, i.e., correctly guessing the state without disturbance using LOCC becomes more difficult. For a set of MESs, the capability of LOCC is completely suppressed, as it cannot surpass random guessing without measurements. This significantly strengthens the restriction on LOCC beyond conventional LD, regardless of the structure of MESs~\cite{nathanson2005distinguishing}.

The limitation of local operations in quantum state discrimination, as depicted by our trade-off relation, can be overcome by utilizing pre-shared entanglement~\cite{cirac2001entangling, bandyopadhyay2009entanglement, bandyopadhyay2015limitations, bandyopadhyay2016entanglement}.
For non-destructive local discrimination (NDLD) of orthogonal MESs,
we introduce an adaptive and fully non-destructive strategy that offers a strict advantage in pre-shared entanglement cost compared to the teleportation-based strategy, widely adopted in conventional LD~\cite{bandyopadhyay2015limitations, bandyopadhyay2021entanglement}.
Our strategy also has a benefit in conventional state discrimination. We demonstrate that for a specific set of MESs, adopting our strategy results in a net entanglement gain after state discrimination that is strictly positive, whereas it cannot be achieved by the conventional teleportation-based method. This provides a novel insight into reusing entanglement in subsequent information processing~\cite{doi:10.1126/sciadv.adi5261}, going beyond the catalytic usage of entanglement with zero gain~\cite{yu2012four}.

As a practical application of our trade-off relation, we construct an entanglement certification protocol, which can be readily extended to test quantum networks with multiple nodes. The proposed scheme can certify entanglement between distant parties even when classical communication is allowed and is robust against depolarizing noise.

\section{Results}
\subsection{Guessing--disturbance trade-off in LD}
\begin{figure}[t]
\includegraphics[width=240px]{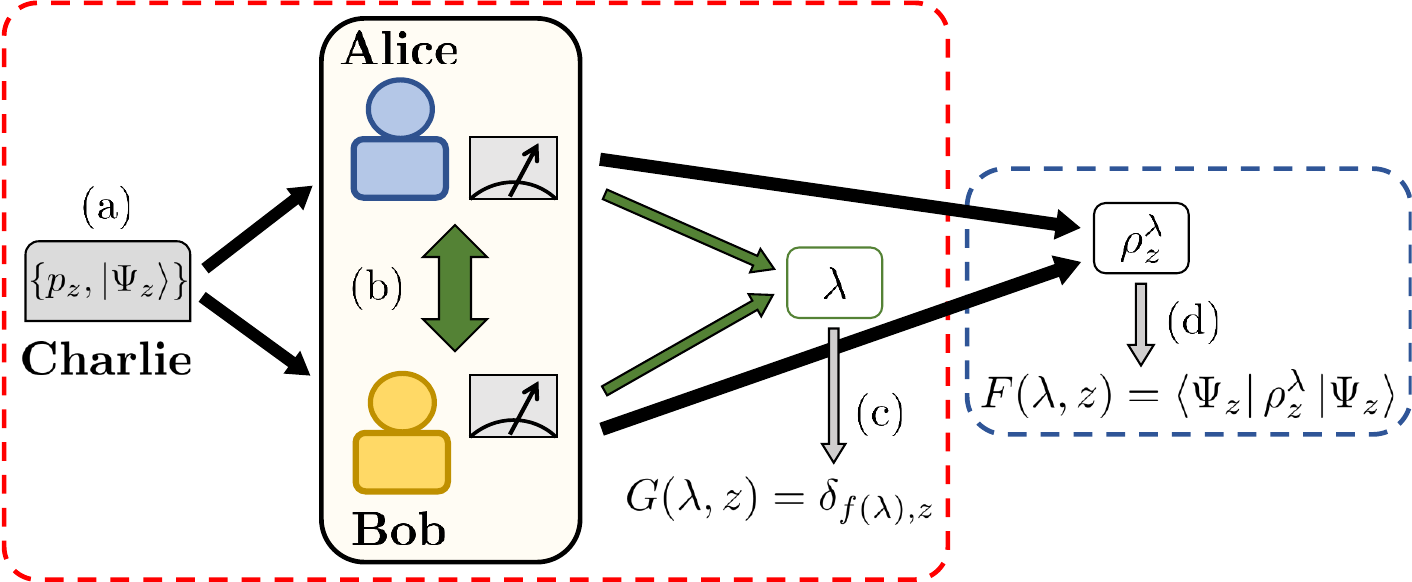}
\caption{Schematic of LD task (red dashed box) and the remaining quantum state (blue dashed box). The black (green) arrow represents quantum (classical) transmission.  (a) Charlie chooses a state from the ensemble $\{ p_z,\ket{\Psi_z}\}$ and sends it to Alice and Bob. (b) Alice and Bob perform LOCC. (c) A classical answer $z_{\text{guess}}=f(\lambda)$ from the local measurement outcomes determines the guessing score $G(\lambda,z)=\delta_{f(\lambda),z}$.  (d) The disturbance of the state can be quantified by the fidelity $F(\lambda,z)=\bra{\Psi_z}\rho^{\lambda}_z\ket{\Psi_z}$.}
\label{fig:NDLD}
\end{figure}
In conventional LD~\cite{barnett2009quantum,bae2015quantum}, the referee Charlie chooses a state from a set of $k$ different quantum states $\{ \ket{\Psi_z} \}_{z=1}^k$ with probability $p_z$ and sends it to two local parties Alice and Bob (see Fig.~\ref{fig:NDLD}). Alice and Bob then guess the state by performing a measurement, described by a positive operator-valued measurement (POVM) $\Pi^{\lambda}$, satisfying $\Pi^{\lambda} \geq 0$ and $\sum_{\lambda=1}^N \Pi^{\lambda}=1$ with $N\geq k$. The probability to get an outcome $\lambda$ for Charlie's choice $z$ is $p(\lambda|z) = \Tr\left[\Pi^{\lambda}\rho_z \right]$ with $\rho_z=\ket{\Psi_z}\bra{\Psi_z}$. For each measurement outcome $\lambda$, Alice and Bob guess the $z$ with some guessing function $f(\lambda) \in \{ 1,2,\cdots, k\}$. The success probability of the correct guess then becomes 
\begin{equation}
p_{\text{succ}}=\overline{G}=\sum_{z,\lambda} p(\lambda,z)G(\lambda,z),
\end{equation}
which can be interpreted as an average of a score function $G(\lambda,z)=\delta_{f(\lambda),z}$ over the joint distribution of $z$ and $\lambda$, $p(\lambda,z) = p_z p(\lambda|z)$.
The success probability heavily depends on Alice and Bob's ability to perform POVMs $\Pi^\lambda $~\cite{bennett1999quantum,ghosh2001distinguishability,yu2014distinguishability,bandyopadhyay2015limitations}. The most generally adopted condition in LD is that Alice and Bob are only allowed to implement measurements that can be realized by LOCC~\cite{barnett2009quantum,bae2015quantum}.

We extend this setup by considering not only the classical answer for guessing but also the quantum state after the measurements. For an outcome $\lambda$, the post-measurement state can be written as $\rho_z^\lambda = \frac{K^\lambda \rho_z K^{\lambda \dagger}}{p(\lambda|z)}$, where $K^\lambda$ are Kraus operators composing the POVMs, $\Pi^{\lambda} = K^{\lambda \dagger} K^\lambda$. For each $\ket{\Psi_z}$, state disturbance after the measurement with outcome $\lambda$ is then captured by the fidelity between the initial state, $F(\lambda,z) = \bra{\Psi_z} \rho^{\lambda}_z \ket{\Psi_z}$. Consequently, the average fidelity after the discrimination can be expressed as $\overline{F} = \sum_{z,\lambda} p(\lambda, z) F(\lambda, z)$.

\begin{figure*}[t]
\includegraphics[width=500px]{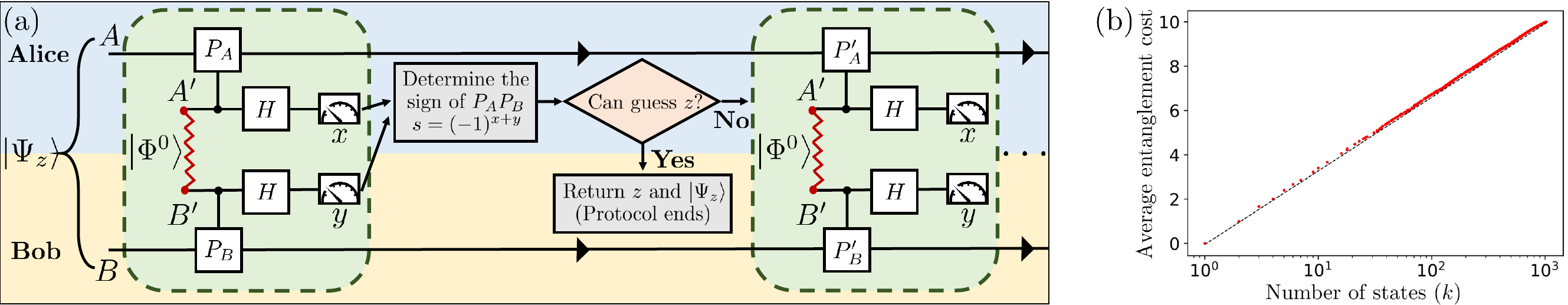}
\caption{(a) Illustration of the adaptive non-destructive scheme. In each round (green-dashed box), controlled-Pauli operations are performed in both local parties using a pre-shared Bell state $\ket{\Phi^0}_{A'B'}$. From the sign information of $P_AP_B$, one can check whether the unambiguous determination of the state is possible so that the protocol ends. Otherwise, the next round of protocol proceeds for the sign of another stabilizer generator $P'_A P'_B$. This procedure continues until $z$ is determined. (b) Average entanglement cost (in ebits) for randomly generated $10$-Bell pair states (local dimension $d = 2^{10}$) as the number of states 
$k$ varies. The cost of our strategy (see Appendix~\ref{app:Entcost}) nearly saturates the lower bound $\log_2k$ (black line).}
\label{fig:scheme}
\end{figure*}

In order to take into account both the correct guessing and disturbance, we introduce the total score
\begin{equation}
    \label{eq:succ_nd1}
    {\cal S} =\overline{G \cdot F}=\sum_{z,\lambda} p(\lambda, z) \delta_{f(\lambda),z}\bra{\Psi_z}\rho^{\lambda}_z\ket{\Psi_z}.
\end{equation}
We note that the maximum value ${\cal S} = 1$ is achieved if and only if the correct guessing is possible without disturbing the state for every $z$.
Notably, we show a tight upper bound on the total score as follows:
\begin{theorem} \label{thm:NDLD_prob}
Suppose that $k$ different bipartite states $\{ \ket{\Psi_z}_{AB} \}_{z=1}^k$ with local dimensions $d_A$ and $d_B$ are distributed with probabilities $p_z$. For any separable operation in the form of $K^\lambda = K_A^\lambda \otimes K_B^\lambda$,
\begin{equation}
{\cal S} = \overline{G \cdot F} \leq d_A d_B \max_{z}\{p_z \xi_z^4\} \equiv {\cal B},
\end{equation}
where $\xi_z$ is the maximum Schmidt coefficient of $\ket{\Psi_z}_{AB}$. Consequently, a trade-off between the average guessing probability and the average fidelity is given as
\begin{equation}
    \overline{G}+\overline{F}\leq 1 + {\cal B}.
\end{equation}
\end{theorem}
The proof of the theorem can be found in Appendix~\ref{App:thm1}. As separable operations strictly include LOCC~\cite{bennett1999quantum,chitambar2012increasing}, the bound holds for any measurement using LOCC. Theorem~\ref{thm:NDLD_prob} directly implies that perfect LD without disturbance is not possible for ${\cal B} < 1$. We highlight that ${\cal B}$ is directly related to the entanglement of the states to be discriminated, given in terms of the Schmidt coefficients~\cite{aniello2009relation, sperling2011schmidt}. In particular, Theorem~\ref{thm:NDLD_prob} leads to a strong restriction on the local discrimination of MESs as follows:
\begin{corollary} \label{thm:NDLD_prob_MES}
For equally distributed $k$ different MESs with the same local dimension $d=d_A=d_B$, the following trade-off relation holds for any separable operation:
\begin{equation}\label{eq:avg_bound}
    {\cal S} \leq \frac{1}{k} \Longrightarrow \overline{G}+\overline{F}\leq 1+\frac{1}{k}.
\end{equation}
\end{corollary}
The trade-off relation states that any local measurement cannot simultaneously achieve perfect guessing and non-destructiveness for discriminating any MESs with $k \geq 2$. In particular, the capability of LOCC is completely suppressed when considering totally non-destructive discrimination, i.e., $\overline F = 1$, in which case $\overline G \leq 1/k$. This can be compared with the upper bound on the guessing probability in conventional state discrimination without considering state disturbance, $\overline{G} \leq d/k$, which scales in the local dimension $d$~\cite{nathanson2005distinguishing}. For example, any two orthogonal MESs ($k=2$) can be perfectly discriminated by local measurements~\cite{walgate2000local}, but our result implies that without disturbing the state, the guessing probability cannot exceed $1/2$. Furthermore, the bound in Eq.~\eqref{eq:avg_bound} tightly suppresses the capability of local measurements, as it can be saturated by random guessing without performing measurements. Our result also fully generalizes the previous studies on quantum delocalized interaction~\cite{paige2020quantum, Bohdan} to be applied to any separable operations.

\subsection{Pre-shared entanglement cost for perfect NDLD}
While Theorem~\ref{thm:NDLD_prob} and Corollary~\ref{thm:NDLD_prob_MES} state that local operations alone cannot perfectly discriminate orthogonal entangled states without disturbance, this limitation can be overcome aided by pre-shared entanglement. The most widely adopted strategy is using quantum teleportation to localize the target state to one of the parties, making perfect discrimination possible. This is proven to be the optimal strategy with the minimum entanglement cost in conventional LD for some cases~\cite{bandyopadhyay2015limitations, bandyopadhyay2021entanglement}. However, once teleportation is performed, the entanglement of the target state also vanishes. Thus, when taking into account the non-destructiveness condition, additional entanglement is required to reprepare the target state via LOCC after its correct guessing. This leads to the question of whether there exists a more efficient strategy than the teleportation to utilize pre-shared entanglement.

In this manuscript, we focus on the pre-shared entanglement cost for achieving perfect NDLD (${\cal S} = 1$) of $k$ different orthogonal MESs. We derive a following lower bound on the entanglement cost:
\begin{theorem}\label{thm:NDLD_ent}
    A perfect NDLD for equally distributed $k$ different orthogonal MESs requires at least $\log_2 k$ ebits of pre-shared entanglement.
\end{theorem}
The lower bound also applies to the average entanglement cost when considering an adaptive strategy with different entanglement costs for each instance (see Appendix~\ref{app:th2}).
We emphasize that the bound is independent in the local dimension $d$. This implies that a na{\"i}ve teleportation of the entire local state might not be the optimal strategy when $d \gg k$, as it costs $2\log_2 d$ ebits including repreparation, resulting in a large gap between the lower bound.

Interestingly, we find an entirely non-destructive and adaptive strategy based on the stabilizer formalism~\cite{gottesman1997stabilizer} that can be more efficient than the teleportation-based scheme in perfect NDLD. As an illustrative example, let us consider Bell states $\ket{\Phi^j}_{AB}$ with $j=0,1,2,3$ in the standard Bell basis.
These four Bell states can be fully characterized by $(s_X, s_Z)$ with $s_X, s_Z \in \{ +1, -1 \}$ based on the following relation:
$$
\begin{aligned}
    X_A X_B \ket{\Phi^j} &= s_X \ket{\Phi^j}, \\
    Z_A Z_B \ket{\Phi^j} &= s_Z \ket{\Phi^j},
\end{aligned}
$$
where $X_{A(B)}$ and $Z_{A(B)}$ are Pauli $X$ and $Z$ operators acting on the subsystem $A(B)$. For example, $\ket{\Phi^0} = (\ket{00} + \ket{11})/\sqrt{2}$ satisfies $X_A X_B \ket{\Phi^0} = (+1) \ket{\Phi^0}$ and $Z_A Z_B \ket{\Phi^0} = (+1) \ket{\Phi^0}$ so that $(s_X, s_Z) = (+1, +1 )$. In this case, we say $\ket{\Phi^0}$ is stabilized by a set of Pauli operators generated by $+X_A X_B$ and $+Z_A Z_B$.
Similarly, the other Bell states are stabilized by a set of Pauli operators generated by $s_X X_A X_B$ and $s_Z Z_A Z_B$ with the different signs $(s_X, s_Z)$. Therefore, if we can determine the signs $s_X$ and $s_Z$ of the generators, we can specify the Bell state. This can be done in a non-destructive manner by using parity measurements introduced in the stabilizer quantum error correction codes~\cite{gottesman1997stabilizer}. For example, $s_Z$ can be determined by the measurement operator $\left\{ \frac{\mathbb{1} \pm Z_{A} Z_{B} }{2} \right\}$.
If the Bell state $\ket{\Phi^j}$ is stabilized by $s_Z Z_A Z_B$, i.e., $Z_A Z_B \ket{\Phi^j} = s_Z \ket{\Phi^j}$, then the probability of each outcome is given by $p_\pm = \bra{\Phi^j} \frac{\mathbb{1} \pm Z_{A} Z_{B} }{2} \ket{\Phi^j} = \frac{1 \pm s_Z}{2}$. This results in a deterministic outcome with the same sign as $s_Z$, while the measurement is non-destructive, as the post-measurement state remains unchanged. The other sign, $s_X$, can be determined in similarly by taking the measurement operator $\left\{ \frac{\mathbb{1} \pm X_{A} X_{B} }{2} \right\}$.

However, such a parity measurement is a nonlocal operator, which cannot be directly implemented via LOCC. Nevertheless, this can be done with the aid of a pre-shared Bell state $\ket{\Phi^0}_{A'B'}$ between the local parties. More precisely, one can apply a controlled unitary operator $U_{P_{A(B)}} = \mathbb{1}_{A(B)} \otimes \ket{0}\bra{0}_{A'(B')} + P_{A(B)} \otimes \ket{1}\bra{1}_{A'(B')}$ for Pauli operators $P_{A(B)} = X_{A(B)}$ or $P_{A(B)} = Z_{A(B)}$ on both $A A'$ and $B B'$ followed by the measurements on the auxiliary systems $A'$ and $B'$ in $\{ \ket{\pm}=(\ket{0}\pm\ket{1})/\sqrt{2} \}$ basis. If both measurement outcomes are the same, the $AB$ system is projected onto $\frac{\mathbb{1} + P_{A}P_{B}}{2}$, otherwise,  $\frac{\mathbb{1} - P_{A}P_{B}}{2}$. Therefore, $1$ ebit is consumed to determine each sign of generators, and the state after the measurement remains unchanged. We continue this procedure until the state is unambiguously deduced (see Fig.~\ref{fig:scheme} (a) and Appendix~\ref{app:parity}). Figure~\ref{fig:scheme} (b) shows that for randomly generated $n$-Bell pairs  $\ket{\Phi^{j_1 \cdots j_n}}_{AB} = \ket{\Phi^{j_1}}_{A_1 B_1} \otimes \cdots \otimes \ket{\Phi^{j_n}}_{A_n B_n}$ with $j_1, \cdots, j_n \in \{ 0,1,2,3\}$, the average entanglement cost of our strategy nearly saturates the lower bound in Theorem~\ref{thm:NDLD_ent}~(see Appendix~\ref{app:Entcost} for the strategy).

To demonstrate a strict advantage of our scheme compared to the teleportation-based one, let us consider perfect NDLD of equally distributed three Bell states, $\{ \ket{\Phi^0}, \ket{\Phi^1}, \ket{\Phi^2} \}$. The signs $( s_X, s_Z )$ distinguishing those Bell states are $\ket{\Phi^0}: ( +, + ), \ket{\Phi^1}:( - , + ), \ket{\Phi^2}: ( +, - ) $. We first determine the sign $s_Z$ by consuming $1$ ebit. When $s_Z = -1$, the state is deduced as $\ket{\Phi^2}$ and
the protocol terminates. When $s_Z = +1$,
we further check $s_X$ to distinguish between $\ket{\Phi^0}$ and $\ket{\Phi^1}$ by consuming additional $1$ ebit (i.e., total $2$ ebits). The average entanglement cost of the overall process then becomes $\frac{1\times 1 + 2 \times 2}{3} = \frac{5}{3}$ ebits. This is strictly less than that of the teleportation-and-repreparation scheme, which costs a total of $2$ ebits ($1$ ebit each for the teleportation and repreparation). Such a gap arises as the stabilizer method can utilize the sign information more compactly by determining each sign adaptively, while the teleportation-based method can only access a pair of signs $(s_X, s_Z)$ at once.

Furthermore, this gap is extendable by considering $n$-Bell pairs $\ket{\Phi^{j_1 \cdots j_n}}_{AB}$
only with three indices for each $j_i \in \{0,1,2\}$. For a set of $n$-Bell pairs $\{ \ket{\Phi^{j_1 \cdots j_n}}_{AB}\}$ with all possible combinations of $(j_1,\cdots,j_n)$'s, i.e., $k=3^n$, the entanglement cost of the stabilizer method is $5n/3$ ebits, while teleportation-and-repreparation of each Bell-pair requires total $2n$ ebits for the perfect NDLD. Hence, the gap $n/3$ grows linearly by increasing the number of Bell pairs $n$ (see Appendix~\ref{app:Entcost}).

Our scheme can readily be generalized to multipartite entangled states. For example, we can show an advantage on the entanglement cost for perfect NDLD of GHZ states compared to the teleportation-based method (see Appendix~\ref{app:GHZ}). 

\subsection{Entanglement earning from NDLD}
We can show our protocol leads to an interesting implication even for conventional LD. This stems from the fact that one can reuse the entanglement during the discrimination process by utilizing the non-destructive nature of the protocol. More explicitly, let us consider the following $6$ MESs,
$\left\{ \ket{\Phi^{012}}, \ket{\Phi^{021}}, \ket{\Phi^{102}}, \ket{\Phi^{120}}, \ket{\Phi^{201}}, \ket{\Phi^{210}} \right\}$.
The average entanglement cost for the perfect NDLD using the stabilizer method is $8/3$ ebits, which is lower than the entanglement (3 ebits) of the target states (see Appendix~\ref{app:Entcost}). In other words, a net gain of entanglement becomes strictly positive after the state discrimination, as one can reuse the Bell pair discriminated in the previous round for the next round of discrimination. This can be interpreted as \textit{entanglement earning from discrimination}, further extending the notion of \textit{entanglement discrimination catalysis} \cite{yu2012four}.

Although finding an exact condition for such phenomenon and computing the optimal entanglement cost are demanding, we note some lower bounds on the entanglement cost. In the above example, a lower bound on the average entanglement cost for the perfect NDLD is $\log_26 \approx 2.58$ ebits from Theorem~\ref{thm:NDLD_ent}, which is strictly smaller than that of the stabilizer method $8/3 \approx 2.67$ ebits.
However, we can show that the latter is the optimal cost under the stabilizer method:
\begin{theorem}\label{th:lower-bound}
    For equally distributed $k$ orthogonal $n$-Bell pairs, a lower bound on the average entanglement cost for the perfect NDLD of the stabilizer method is $m+2-\frac{2^{m+1}}{k}$ ebits, 
    where $m=\lfloor\log_2k\rfloor$.
\end{theorem}
Putting $k=6~(m=2)$, the cost of our strategy saturates the lower bound, 8/3 ebits. A proof of theorem~\ref{th:lower-bound} and comparison with a teleportation-based strategy are given in Appendix.~\ref{app:Entcost}.

\begin{figure}[t]
\includegraphics[width=240px]{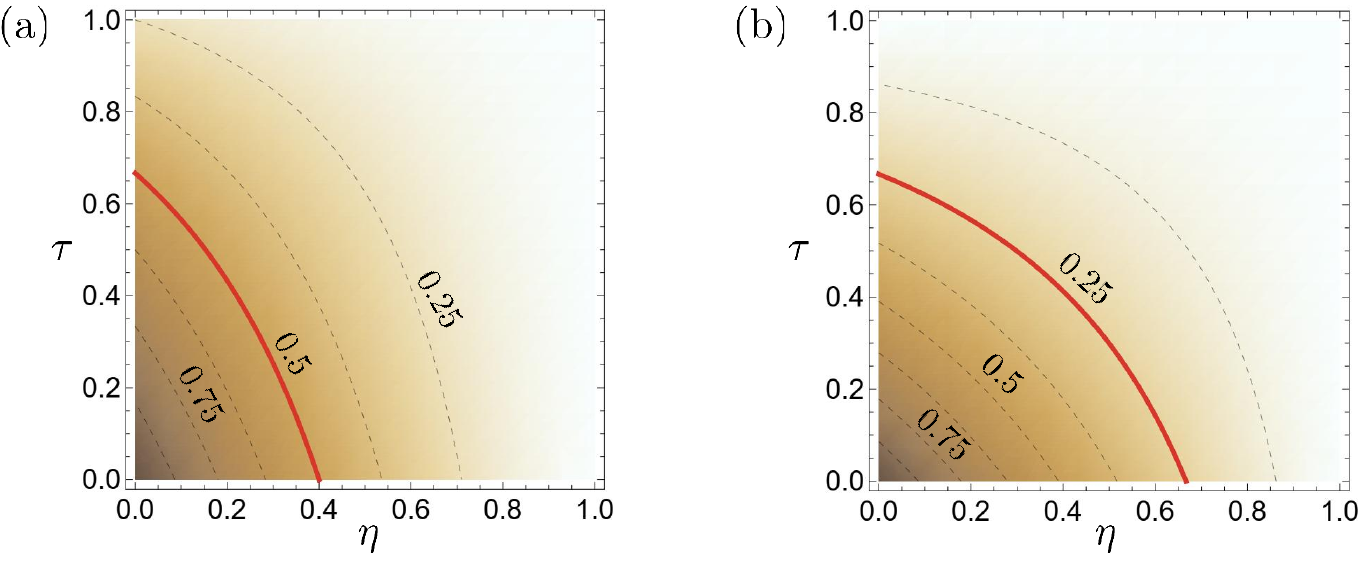}
\caption{Achievable total score for entanglement certification protocol using $k$ different Bell states in noisy cases. (a) $k=2$, (b) $k=4$ with noise parameters $\eta$ and $\tau$ for the initial state and pre-shared entanglement, respectively. Certification successes within regions inside red lines, where the total score is larger than $1/k$. }
\label{fig:werner}
\end{figure}
\subsection{Application to entanglement certification}
As a practical application, we discuss how our scheme can be utilized in entanglement certification. To do this, the referee, Charlie, first chooses one state among $k$ orthogonal MESs, which can be done by performing a local unitary operation on any given MES.
Charlie then distributes the state to Alice and Bob. After the discrimination, each party returns the state to Charlie of the same dimension as the state they originally received. Then Charlie performs a projection measurement onto the initial state to evaluate the total score ${\cal S}$. When ${\cal S} > 1/k$, entanglement between Alice and Bob is certified, as such a score cannot be achieved by LOCC from Corollary~\ref{thm:NDLD_prob_MES}. While transmission of entangled quantum states required in our protocol could be more demanding, it has some distinct advantages compared to other protocols based on nonlocality tests~\cite{shalm2015strong,bowles2018device} that classical communication is fully allowed between local parties.

In a realistic situation, however, noisy communication channels and imperfect pre-shared entanglement can degrade the total score. 
Nevertheless, for a local noise channel, the bound ${\cal S} > 1/k$ remains valid, as any local strategy including the local noise channel can still be represented by separable Kraus operators.
Specifically, let us consider a situation where depolarizing noises, i.e., ${\cal E}_\eta(\rho) = (1-\eta)\rho+\eta\mathbb{1}/d^2$ with $\eta \in [0,1]$, cause the MESs to evolve into Werner-like states $\rho_\eta^i = {\cal E}_\eta(\ket{\Phi^i}\bra{\Phi^i}) = (1-\eta) \ket{\Phi^i}\bra{\Phi^i}+\eta\mathbb{1}/d^2$. Achievable total score for the noisy channel ${\cal E}_\eta$ and noisy pre-shared entangled state $\rho^0_{\tau}$ depicted in Fig.~\ref{fig:werner} shows the robustness of our scheme within a moderate range of parameters (for more details, see Appendix~\ref{app:cert}). We note that when $k=d^2$, our protocol operates as long as the states evolved under the noise channel are entangled, i.e., $\eta < d/(d+1)$. Conversely, when the channel is noiseless ($\eta = 0$), our protocol can certify the entanglement of any entangled Werner-like states with $\tau < d/(d+1)$, which cannot be done by nonlocality tests~\cite{werner1989quantum, barret2002nonsequential, almeida2007noise, oszmaniec2017simulating, hirsch2017betterlocalhidden}.

We extend this approach to multipartite scenarios. To certify entanglement between three nodes in a quantum network, the referee tests NDLD for $k$ different GHZ states $(1/\sqrt{2}) (\ket{x_1 x_2 x_3} \pm \ket{\bar{x}_1 \bar{x}_2 \bar{x}_3})$, where $x_i \in \{ 0, 1 \}$ and $\bar{x} = x + 1~({\rm mod}~2)$. By noting that any multipartite LOCC is a subset of LOCC between the bipartition of the total system and any bipartition of GHZ states is maximally entangled with uniform Schmidt coefficients $1/\sqrt{2}$, the total score using LOCC is bounded above by ${\cal S} \leq 2/k$ by Theorem~\ref{thm:NDLD_prob}. Hence, choosing any three GHZ states is sufficient to certify entanglement by witnessing the total score higher than $2/3$. In contrast, without the state disturbance condition, some sets of up to four GHZ states can be perfectly discriminated by using LOCC~\cite{hayashi2006bounds}.

For a general $N$-node quantum network, we can adopt the absolutely maximally entangled (AME) states~\cite{helwig2012absolute, helwig2013absolutely, goyeneche2015absolutely}, any bipartition of which leads to an MES. In particular, when $N$ is even, the highest total score for $k$ different AME states becomes $1/k$, which is a multiparty extension of Corollary~\ref{thm:NDLD_prob_MES}. This is because any multipartite local strategy for AME states is a subset of a bipartite local strategy for MESs with the same local dimension by taking the $N/2$ node bipartition. We note that AME states exist for any number of parties with sufficiently high local dimension~\cite{helwig2013absolutely} (see also, Refs.~\cite{scott2004multipartite,huber2018bounds} for limitations in low dimensions) and can be systematically constructed~\cite{helwig2013absolutely, raissi2018optimal, raissi2020constructions, raissi2022general}.  
Alternatively, we can exploit randomly generated stabilizer states, which approximately satisfy the AME property with high probability~\cite{page1993average,zyczkowski2001induced} so that the bound on the total score slightly deviates from that of random guessing (see Appendix~\ref{app:random}).

\section{Remarks}
We have established a trade-off relation between guessing and disturbance in LD of entangled states. This imposes a strong restriction on LOCC, as it cannot perform better than random guessing for MESs, regardless of their choice. We have also presented a non-destructive and adaptive strategy for efficiently utilizing pre-shared entanglement to achieve perfect NDLD, surpassing the conventional strategy based on teleportation. Our results also provide practical applications to conventional state discrimination and entanglement certification. 

From a thermodynamic perspective, the advantage of the proposed non-destructive strategy lies in overcoming the inherent irreversibility between measurement and repreparation of quantum states. In this direction, constructing a general formalism that precisely describes the difference between destructive and non-destructive strategies in terms of irreversibility will be an interesting future work. A possible approach is to analyze additional entanglement cost for the initial state repreparation, based on the irreversibility of entanglement concentration~\cite{kumagai2013entanglement}.

\begin{acknowledgments}
This work was supported by Institute of Information \& Communications Technology Planning \& Evaluation (IITP) grant funded by the Korea government (MSIT) (No. 2022-0-00463). M.H. and H.K. are supported by the KIAS Individual Grant Nos. QP089801 (M.H.) and CG085301 (H.K) at Korea Institute for Advanced Study.
\end{acknowledgments}

\onecolumngrid
\appendix

\section{Proof of Theorem 1}\label{App:thm1}
\begin{proof}
Let us consider a set of $k$ different bipartite pure states $\{ \ket{\Psi_z}_{AB} \}_{z=1}^k$ with associated probability $p_z$. If Alice and Bob can only perform a separable measurement, described by Kraus operators $K^{\lambda}_{AB}=K^{\lambda}_A \otimes K^{\lambda}_B$, the initial state transforms into $\rho_z^\lambda = \frac{K_{AB}^\lambda \rho_z K_{AB}^{\lambda \dagger}}{p(\lambda|z)}$ for an outcome $\lambda$. An upper bound on the total score ${\cal S} = \overline{G \cdot F} = \sum_{\lambda,z} p(\lambda, z) G(\lambda, z) F(\lambda,z)$ with $p(\lambda,z) = p(\lambda|z) p_z$, $G(\lambda, z) = \delta_{f(\lambda),z}$ and $F(\lambda,z) = \bra{\Psi_z} \rho^\lambda_z \ket{\Psi_z}$ can be found as 
\begin{align*}
    {\cal S}
    &=\sum_{z=1}^k \sum_{\lambda} p_z  p(\lambda|z) G(\lambda, z) F(\lambda, z)\\ 
    &=\sum_{z=1}^k p_z \sum_{\lambda} p(\lambda|z) \delta_{f(\lambda),z}\bra{\Psi_z}\rho^{\lambda}_z\ket{\Psi_z}\\ 
    &=\sum_{z=1}^k p_z \sum_{\lambda_z} \left|\bra{\Psi_z}K^{\lambda_z}_A\otimes K^{\lambda_z}_B \ket{\Psi_z}\right|^2\\
    &=\sum_{z=1}^k p_z \sum_{\lambda_z}\Tr\left[ \ket{\Psi_z}\bra{\Psi_z}K^{\lambda_z \dagger}_A \otimes K^{\lambda_z \dagger}_B \ket{\Psi_z}\bra{\Psi_z}K^{\lambda_z}_A \otimes K^{\lambda_z}_B\right]\\
    &=\sum_{z=1}^k p_z \sum_{\lambda_z}\Tr\left[ \ket{\Psi_z}\bra{\Psi_z}(K^{\lambda_z \dagger}_A \otimes \mathbb{1}_B)(\mathbb{1}_A \otimes K^{\lambda_z \dagger}_B) \ket{\Psi_z}\bra{\Psi_z}(K^{\lambda_z}_A \otimes \mathbb{1}_B)(\mathbb{1}_A \otimes K^{\lambda_z}_B)\right]\\
    &=\sum_{z=1}^k p_z \sum_{\lambda_z}\Tr\left[ \ket{\Psi_z}\bra{\Psi_z}(\mathbb{1}_A \otimes K^{\lambda_z \dagger}_B)(K^{\lambda_z \dagger}_A \otimes \mathbb{1}_B) \ket{\Psi_z}\bra{\Psi_z}(K^{\lambda_z}_A \otimes \mathbb{1}_B)(\mathbb{1}_A \otimes K^{\lambda_z}_B)\right]\\
    &=\sum_{z=1}^k p_z \sum_{\lambda_z}\Tr\left[ (K^{\lambda_z \dagger}_A \otimes \mathbb{1}_B) \ket{\Psi_z}\bra{\Psi_z}(K^{\lambda_z}_A \otimes \mathbb{1}_B)(\mathbb{1}_A \otimes K^{\lambda_z}_B)\ket{\Psi_z}\bra{\Psi_z}(\mathbb{1}_A \otimes K^{\lambda_z \dagger}_B)\right]\\
    &\leq \sum_{z=1}^k p_z \sum_{\lambda_z} \Tr_A\left[ K^{\lambda_z}_A K^{\lambda_z \dagger}_A \Tr_B(\ket{\Psi_z}\bra{\Psi_z})\right]\Tr_B\left[ K^{\lambda_z \dagger}_B K^{\lambda_z }_B\Tr_A(\ket{\Psi_z}\bra{\Psi_z})\right]\\
    &= \sum_{z=1}^k p_z \sum_{\lambda_z}\Tr_A\left[ K^{\lambda_z}_A K^{\lambda_z \dagger}_A \rho_{A,z} \right]\Tr_B\left[ K^{\lambda_z \dagger}_B K^{\lambda_z }_B \rho_{B,z}\right],
\end{align*}
where $\rho_{A,z}$ and $\rho_{B,z}$ are reduced density matrices of $\ket{\Psi_z}_{AB}\bra{\Psi_z}$. From the Schmidt decomposition $\ket{\Psi_z}_{AB}= \sum_{i=1}^d r_i^z \ket{\psi_i^z}_A \otimes \ket{\psi_i^z}_B$, where $d={\min(d_A,d_B)}$ and $d_A$ and $d_B$ are the dimensions of the local systems $A$ and $B$, one can obtain the eigendecomposition of the local states $\rho_{A,z}= \sum_{i=1}^d (r_i^z)^2 \ket{\psi_i^z}_A\bra{\psi_i^z}$ and $\rho_{B,z}= \sum_{i=1}^d (r_i^z)^2 \ket{\psi_i^z}_B \bra{\psi_i^z}$. By noting that $\rho_{A,z} \leq \xi_z^2 \mathbb{1}_A$ and $\rho_{B,z} \leq \xi_z^2 \mathbb{1}_B$ with $\xi_z = \max \{ r_i^z \}$, we have a bound on the total score,
\begin{align*}
    {\cal S} &\leq \sum_{z=1}^k p_z \sum_{\lambda_z}\Tr_A\left[ K^{\lambda_z}_A K^{\lambda_z \dagger}_A \rho_{A,z} \right]\Tr_B\left[ K^{\lambda_z \dagger}_B K^{\lambda_z }_B \rho_{B,z}\right]\\
    &\leq \sum_{z=1}^k p_z \xi_z^4 \sum_{\lambda_z}\Tr_A\left[ K^{\lambda_z}_A K^{\lambda_z \dagger}_A \right]\Tr_B\left[ K^{\lambda_z \dagger}_B K^{\lambda_z }_B \right]\\
    &\leq \left( \max_z\{p_z \xi_z^4\} \right) \sum_{z=1}^k \sum_{\lambda_z}\Tr_A\left[ K^{\lambda_z}_A K^{\lambda_z \dagger}_A \right]\Tr_B\left[ K^{\lambda_z \dagger}_B K^{\lambda_z }_B \right]\\
    &= \left( \max_z\{p_z \xi_z^4\} \right) \sum_{\lambda}\Tr \left[ \left( K^{\lambda \dagger}_A \otimes K^{\lambda \dagger}_B \right) \left( K^{\lambda}_A \otimes K^{\lambda }_B \right) \right]\\
    &= \left( \max_z\{p_z \xi_z^4\} \right) \Tr \left[\mathbb{1}_A \otimes \mathbb{1}_B \right]\\
    &= d_A d_B \max_z\{p_z \xi_z^4\}\equiv {\cal B}.
\end{align*}
To show the trade-off relation, Eq.~(4) in the main text, let us consider inequalities for given $\lambda$ and $z$,
\begin{equation*}
   G(\lambda, z) + F(\lambda, z) \leq 1 + G(\lambda, z) \cdot F(\lambda, z),
\end{equation*}
for $0 \leq G(\lambda, z) \leq 1$ and $0 \leq F(\lambda, z) \leq 1$. After taking the average over $p(\lambda,z)$, we finally get the trade-off as
\begin{align*}
    \overline{G}+\overline{F} &= \sum_{\lambda,z} p(\lambda,z) [G(\lambda,z) + F(\lambda,z)] \\
    &\leq \sum_{\lambda,z} p(\lambda,z) [1 + G(\lambda,z) \cdot F(\lambda,z)] \\
    &=1 + \overline{G \cdot F} \\
    &\leq 1+{\cal B},
\end{align*}

where the second inequality comes from $\overline{G \cdot F}={\cal S}\leq {\cal B}$.
\end{proof}

\section{Proof of Theorem 2}\label{app:th2}

\begin{proof}
Suppose that Alice and Bob have a pre-shared MES $\ket{\alpha}=1/\sqrt{f}\sum_{\ell=1}^f \ket{\ell}_{A'} \otimes \ket{\ell}_{B'}$. We write the given MES $\ket{\Psi_z}$ by $1/\sqrt{d}\sum_{i=1}^d \ket{\psi_i^z\psi_i^z}_{AB}$ for each $z$ by appropriately choosing the basis of Alice and Bob.
Since the initial state is changed to $\ket{\Psi_z}\otimes \ket{\alpha}$, the total score is given by
\begin{align*}
    {\cal S}&=\frac 1 k \sum_z \sum_{\lambda_z} \left\|\bra{\Psi_z}_{AB} K^{\lambda_z}_{AA'} \otimes K^{\lambda_z}_{BB'} \ket{\Psi_z}_{AB} \ket{\alpha}_{A'B'}\right\|^2
    \\
    &=\frac 1 k \sum_z \sum_{\lambda_z} 
    \sum_{m,n}\left|\bra{\Psi_z}_{AB}\bra{mn}_{A'B'} K^{\lambda_z}_{AA'} \otimes K^{\lambda_z}_{BB'} \ket{\Psi_z}_{AB}\ket{\alpha}_{A'B'}\right|^2
    \\
    &=\frac 1 {d^2 f k} \sum_z \sum_{\lambda_z} 
    \sum_{mn}\left|\sum_{i,j,\ell}\bra{\psi_i^z\psi_i^z}_{AB}\bra{mn}_{A'B'} K^{\lambda_z}_{AA'} \otimes K^{\lambda_z}_{BB'} \ket{\psi_j^z\psi_j^z}_{AB}\ket{\ell\ell}_{A'B'}\right|^2
    \\
    &=\frac 1 {d^2 f k} \sum_z \sum_{\lambda_z} 
    \sum_{m,n}\left|\sum_{i,j,\ell}\bra{\psi_i^zm}_{AA'} K^{\lambda_z}_{AA'}\ket{\psi_j^z\ell}_{AA'} \cdot \bra{\psi_i^zn}_{BB'}K^{\lambda_z}_{BB'} \ket{\psi_j^z\ell}_{BB'}\right|^2
    \\
    &\le \frac 1 {d^2 f k} \sum_z \sum_{\lambda_z} 
    \sum_{m,n}
    \left(\sum_{i,j,\ell}\left|\bra{\psi_i^zm}_{AA'} K^{\lambda_z}_{AA'}\ket{\psi_j^z\ell}_{AA'}\right|^2 \right)\cdot 
    \left(\sum_{i,j,\ell}\left| \bra{\psi_i^zn}_{BB'}K^{\lambda_z}_{BB'} \ket{\psi_j^z\ell}_{BB'}\right|^2\right)
    \\
    &=\frac 1 {d^2 f k} \sum_z \sum_{\lambda_z} 
    \sum_{m,n}
    \Tr_A\left[\bra{m}_{A'} K^{\lambda_z}_{AA'}K^{\lambda_z\dagger}_{AA'}\ket{m}_{A'}\right]\cdot 
    \Tr_B\left[\bra{n}_{B'} K^{\lambda_z}_{BB'}K^{\lambda_z\dagger}_{BB'}\ket{n}_{B'}\right]
    \\
    &=\frac 1 {d^2 f k} \sum_z \sum_{\lambda_z} 
    \Tr_{AA'}\left[ K^{\lambda_z}_{AA'}K^{\lambda_z\dagger}_{AA'}\right]\cdot 
    \Tr_{BB'}\left[ K^{\lambda_z}_{BB'}K^{\lambda_z\dagger}_{BB'}\right]
    \\
    &=\frac 1 {d^2 f k} \sum_z \sum_{\lambda_z} 
    \Tr\left[ 
    (K^{\lambda_z\dagger}_{AA'}\otimes K^{\lambda_z\dagger}_{BB'})
    (K^{\lambda_z}_{AA'}\otimes K^{\lambda_z}_{BB'})
    \right]
    \\
    &=  \frac 1 {d^2 f k}  \Tr\left[\mathbb{1}_{AA'} \otimes \mathbb{1}_{BB'}\right] = \frac{(df)^2}{d^2fk}=  \frac f { k}.
\end{align*}
For the perfect non-destructive local state discrimination (NDLD), $f\geq k$. Therefore, a lower bound on the pre-shared entanglement cost is $\log_2 k$ ebits.
\end{proof}

We can also prove Theorem 2 by using another way even allowing an adaptive strategy with different entanglement costs for each instance. To show this, let us consider the state merging bound for the equally distributed orthogonal $k$ MESs, which is given by the conditional entropy such as~\cite{horodecki2005partial,horodecki2007quantum} 
\begin{equation*}
    S(A|B)_{\rho_z}=S(\rho_z^{AB})-S(\rho_z^B)=\log_2 k- \log_2 d.
\end{equation*}
This means we need at least $\log_2 k-\log_2 d$ ebits for the state merging from system $A$ to $B$ asymptotically. Suppose we perform the state merging using our the perfect NDLD by consuming $\log_2 f$ ebits of pre-shared entanglement. After finishing the NDLD, the remaining state is an MES with $\log_2 d$ ebits. Since the state merging bound of this remaining state is $-\log_2 d$, the total entanglement cost of this state merging protocol is $\log_2 f-\log_2 d$ ebits. Suppose $f<k$. This is not possible even on average because if it were true, the state merging bound would be asymptotically violated by repeating the same protocol. Thus it should be $f \geq k$, which claims the result.

\section{Non-local parity measurements on stabilizer states}\label{app:parity}

Let us assume the states to be distinguished are stabilizer states. When local parties pre-share entanglement $\ket{\Phi^0}_{A'B'}=1/\sqrt{2}(\ket{00}_{A'B'}+\ket{11}_{A'B'})$, the state after the controlled local unitary operation $U_{P_{A(B)}} = \mathbb{1}_{A(B)} \otimes \ket{0}\bra{0}_{A'(B')} + P_{A(B)} \otimes \ket{1}\bra{1}_{A'(B')}$ is given by
\begin{equation}
    \ket{\Psi_z}_{AA'BB'}\equiv U_{P_{A}} \otimes U_{P_{B}} \ket{\Psi_z}_{AB}  \ket{\Phi^0}_{A'B'}=\frac{1}{\sqrt{2}}\left( \ket{\Psi_z}_{AB}\ket{00}_{A'B'}+P_AP_B\ket{\Psi_z}_{AB}\ket{11}_{A'B'}\right). 
\end{equation}
Then we perform measurements in the computational basis on $A'$ and $B'$ after applying Hadamard gates. The corresponding projection operators can be written as $\Pi_{A'B'}^{(x,y)} = H^{\otimes 2} \left( \Pi^x_{A'} \otimes \Pi^y_{B'} \right) H^{\otimes 2}$, where $\Pi^{x}_{A'} = \ket{x}_{A'}\bra{x}$ and $\Pi^{y}_{B'} = \ket{y}_{B'}\bra{y}$ are projection operators in local parties in the computational basis. This also can be explicitly written in the $\{ \ket{\pm}=1/\sqrt{2}(\ket{0}\pm\ket{1})\}$ basis as

\begin{align*}
\Pi^{(0,0)}_{A'B'}&=\ket{+}_{A'}\bra{+}\otimes \ket{+}_{B'}\bra{+},\\ \Pi^{(0,1)}_{A'B'}&=\ket{+}_{A'}\bra{+}\otimes \ket{-}_{B'}\bra{-},\\ \Pi^{(1,0)}_{A'B'}&=\ket{-}_{A'}\bra{-}\otimes \ket{+}_{B'}\bra{+},\\ \Pi^{(1,1)}_{A'B'}&=\ket{-}_{A'}\bra{-}\otimes \ket{-}_{B'}\bra{-}.
\end{align*} 
Consequently, the state after the local measurements for a pair of outcomes $(x,y)$ with $x,y \in \{ 0,1\}$ is given by
\begin{align}
\ket{\Psi_z}_{AA'BB'}\xrightarrow{\Pi^{(x,y)}_{A'B'}}\frac{\mathbb{1}_{AB}+(-1)^{x+y}P_AP_B}{2} \ket{\Psi_z}_{AB}.
\end{align}

Since $ P_A P_B$ is one of generators of stabilizers for $\ket{\Psi_z}_{AB}$ up to sign, $P_A P_B \ket{\Psi_z}_{AB}=s_P \ket{\Psi_z}_{AB}$ with $s_P \in \{ +1,-1\}$. For a given stabilizer generator $P_A P_B$, we can determine the sign as $s_p=(-1)^{x+y}$: if the local measurement outcomes are the same, $s_P=+1$, and if not, $s_P=-1$ with the unit probability. Note that the state after this parity measurement is undisturbed, so we can perform another parity measurement for $P_A' P_B'$ on the same state until we unambiguously determine the state. For example, the stabilizer generators of Bell states are $\{ s_X X_A X_B, s_Z Z_A Z_B \}$ with the signs $s_X, s_Z \in \{ +1, -1 \}$, where $X_{A(B)}$ and $Z_{A(B)}$ denote the Pauli $X$ and $Z$ operators acting on the subsystem $A(B)$. Hence, any Bell state can be perfectly discriminated by reading the signs $s_X$ and $s_Z$ through the parity measurements.

\section{Average entanglement cost of adaptive strategy}\label{app:Entcost}

\subsection{General strategy}
In the previous section, we show the non-local parity measurement $\{ \frac{\mathbb{1}\pm P_A P_B}{2}\}$ can be performed on the stabilizer state $\ket{\Psi_z}$ by using 1 ebit of pre-shared entanglement, $\ket{\Phi^0}$. Here, we introduce an adaptive and non-destructive strategy for the perfect NDLD of a set of stabilizer states, particularly focusing on the $k$ different $n$-Bell pairs. Then we compute the average cost of pre-shared entanglement and compare it with that of the teleportation-based strategy.

An $n$-Bell pairs has $2n$ stabilizer generators, 
$\{ s_{Z_1}Z_{A_1}Z_{B_1},s_{X_1}X_{A_1}X_{B_1},\cdots,s_{Z_n}Z_{A_n}Z_{B_n},s_{X_n}X_{A_n}X_{B_n}\}$, with the signs $s_{Z_i}, s_{X_i} \in \{+1,-1\}$. In other words, for a given set of $k$ $n$-Bell pairs, we have $k$ of $2n$-tuple $(s_{Z_1},s_{X_1},\cdots,s_{Z_n},s_{X_n})$, in which each tuple uniquely represents each state in the set. This can be described by a $k \times 2n$ sign table. For example, for $n=2$ and $k=5$, the sign table is given by
\begin{center}
\setlength{\tabcolsep}{3pt}
\renewcommand{\arraystretch}{1.2}
\begin{tabular}{ c|c|c|c|c } 
  & $s_{Z_1}$  & $s_{X_1}$ & $s_{Z_2}$ & $s_{X_2}$\\ 
 \hline
 $\ket{\Phi^{00}}$ & $+$ & $+$ & $+$ & $+$  \\ 
 $\ket{\Phi^{02}}$ & $+$ & $+$ & $-$ & $+$  \\ 
 $\ket{\Phi^{10}}$ & $+$ & $-$ & $+$ & $+$  \\
 $\ket{\Phi^{13}}$ & $+$ & $-$ & $-$ & $-$   \\
 $\ket{\Phi^{23}}$ & $-$ & $+$ & $-$ & $-$   \\

\end{tabular}
\end{center}
where $\ket{\Phi^{j_1 \cdots j_n}}_{AB} = \ket{\Phi^{j_1}}_{A_1 B_1} \otimes \cdots \otimes \ket{\Phi^{j_n}}_{A_n B_n}$ with $j_1, \cdots j_n \in \{ 0,1,2,3 \}$ in the standard Bell basis and denoting `$+$'$: +1$ and `$-$'$: -1$ for simplicity. For a given sign table, the procedure for the perfect NDLD is as follows: i) Choose a column and read the sign of the corresponding generator by applying the non-local parity measurement. ii) Based on the result of the sign, choose the next column for the sign of another generator. This procedure continues until we collect enough signs to unambiguously determine the state among the set of states. The problem is how to minimize the number of parity measurements, lowering the entanglement cost.

Our strategy is to choose a column with the most balanced signs at each step. Based on the result of the sign measurement at the chosen column, we delete the rows with the opposite sign to the measured one. We then choose the next column among the remaining columns, and continue this process until we can specify the single row for the target state. If multiple columns have the same sign balance at any step, we randomly select one. 

In the above example, the first column ($s_{Z_1}$) has 4 `$+$'s and 1 `$-$', and all other columns have 3 `$+$'s and 2 `$-$'s or 2 `$+$'s and 3 `$-$'s (denoting $(1:4)$ for the first column and $(2:3)$ for others). Since $(2:3)$ is more balanced than $(1:4)$, we can randomly choose a column except the first column. Suppose we select the second column and read the sign $s_{X_1}$. If we get `$+$', whose probability is $3/5$, the target state is one of three candidates, $\{ \ket{\Phi^{00}}, \ket{\Phi^{02}}, \ket{\Phi^{23}} \}$. To discriminate them further, we can choose the next column as the first column $(s_{Z_1})$ because all remaining columns have the same sign balance ($1:2$) after deleting the 3rd and 4th rows. If we get `$-$' for the first column, whose probability is $1/3$, we finally deduce the state as $\ket{\Phi^{23}}$. Otherwise, we search the next column with the most balanced signs, resulting in the 3rd one, $s_{Z_2}$. According to the result of this sign, the final state is either $\ket{\Phi^{00}}$ or $\ket{\Phi^{02}}$. If the sign of the first chosen column $(s_{X_1})$ is `$-$', whose probability is 2/5, we can similarly deduce the target state as either $\ket{\Phi^{10}}$ or $\ket{\Phi^{13}}$. 

Then we can compute the average entanglement cost of pre-shared entanglement by counting the average number of sign readings. In the above example, this can be written as  
\begin{equation*}
    \frac{3}{5}\times \left( \frac{1}{3}\times 2+\frac{2}{3}\times 3\right)+\frac{2}{5}\times 2=\frac{12}{5}~~\text{(ebits)}.
\end{equation*}
In the following, we apply our strategy to various examples in the main text and compute the average entanglement cost.

\subsection{Example: Bell states}
Since the case of two Bell states is trivial, let us consider a set of three Bell states  $\{\ket{\Phi^0}, \ket{\Phi^1}, \ket{\Phi^2}\}$ chosen without loss of generality. The corresponding sign table is given by
\begin{center}
\setlength{\tabcolsep}{3pt}
\renewcommand{\arraystretch}{1.2}
\begin{tabular}{ c|c|c } 
  & $s_Z$ & $s_X$ \\ 
 \hline
 $\ket{\Phi^{0}}$  & $+$ & $+$ \\ 
 $\ket{\Phi^{1}}$ & $+$ & $-$ \\ 
 $\ket{\Phi^{2}}$ & $-$ & $+$  \\
\end{tabular}
\end{center}
We can choose any column first since two columns have the same sign balance ($1:2$). Suppose we select the first column. If we obtain the `$-$' sign for $s_Z$ with the probability 1/3, we can determine the state as $\ket{\Phi^2}$. If we get `$+$' sign with the probability 2/3, we need to check the sign $s_X$ by consuming another ebit. We can then  finally deduce the state as $\ket{\Phi^0}$ for the `$+$' sign and $\ket{\Phi^1}$ for the `$-$' sign with equal probabilities. Hence, the average entanglement cost is $\frac{1\times1+2\times2}{3}=\frac{5}{3}$ ebits. In contrast, the optimal entanglement cost for the conventional state discrimination of three Bell states is 1 ebit~\cite{bandyopadhyay2015limitations}, which can be achieved by performing teleportation. Therefore, the teleportation-and-repreparation strategy costs $1+1=2$ ebits, resulting in a gap of 1/3 ebits compared with our strategy. 

For four Bell states, the sign table is given by
\begin{center}
\setlength{\tabcolsep}{3pt}
\renewcommand{\arraystretch}{1.2}
\begin{tabular}{ c|c|c } 
  & $s_Z$ & $s_X$ \\ 
 \hline
 $\ket{\Phi^{0}}$  & $+$ & $+$ \\ 
 $\ket{\Phi^{1}}$ & $+$ & $-$ \\ 
 $\ket{\Phi^{2}}$ & $-$ & $+$  \\
 $\ket{\Phi^{3}}$ & $-$ & $-$  \\
\end{tabular}
\end{center}
In this case, we can also choose any column first because of the same sign balance ($2:2$). If we select the first column, each sign is obtained with the same probability. For any case, we need to check the sign $s_X$, which also results in each sign with the same probability. Therefore, the average entanglement cost is $\frac{2\times2+2\times2}{4}=2$ ebits, which is the same as the cost of the teleportation-and-repreparation strategy and saturates the lower bound.

\subsection{Example: $n$-Bell pairs with three indices}
Consider $n$-Bell pairs with only three indices:
$\ket{\Phi^{j_1 \cdots j_n}}_{AB}$ with $j_1, \cdots j_n \in \{ 0,1,2\}$. For a set of $n$-Bell pairs $\{ \ket{\Phi^{j_1 \cdots j_n}}_{AB}\}$ with all possible combinations of $(j_1,\cdots,j_n)$'s, i.e., $k=3^n$, we have a $3^n \times 2n$ sign table, where all columns have the sign balance $(1:2)$. Suppose we choose the first column, $s_{Z_1}$. Then we get `$+$' with the probability 2/3 and `$-$' with the probability 1/3, the same with the three Bell states case. If we get `$-$', the first Bell pair is deduced as $\ket{\Phi^2}$, and we start over for $(n-1)$-Bell pairs. If we get `$+$', we should choose the next column as $s_{X_1}$, because it has only the balanced signs, ($1:1$), but all others have ($1:2$). This implies that our strategy for this case involves successively applying the same strategy for the three Bell states to each Bell pair a total of $n$ times, resulting in the average entanglement cost of $5n/3$ ebits.

In the teleportation-and-repreparation, we teleport all $n$-Bell pairs to localize them in one party, which costs $n$ ebits. By considering the repreparation cost of $n$-Bell pairs, it requires a total $2n$ ebits for the the perfect NDLD. Hence, the gap $n/3$ grows linearly by increasing the number of Bell pairs $n$.

\subsection{Example: Random $n$-Bell pairs}

To examine the efficiency of our strategy in a more generic situation, we apply the strategy on randomly generated $n$-Bell pairs, with $k \in [2,2^n]$, and compute the average entanglement cost. We can numerically demonstrate that the average entanglement cost of this case nearly saturates the lower bound $\log_2 k$ ebits, when $n=10$ (Fig.~2 (b) in the main text).

\subsection{Example: 4 MESs (Entanglement discrimination catalysis)}

The previous examples show the efficiency of our strategy for the the perfect NDLD. A natural question is: Can we obtain any interesting result in conventional state discrimination by exploiting our method? 

To answer this, first, we consider conventional local state discrimination for a set of 4 MESs as $\{\ket{\Phi^{00}},\ket{\Phi^{11}},\ket{\Phi^{21}},\ket{\Phi^{31}}\}$, introduced in Ref.~\cite{yu2012four}. Since this set cannot be perfectly discriminated with LOCC~\cite{yu2012four}, we have to use pre-shared entanglement to reach the perfect guessing. A simple method uses quantum teleportation on the first qubit by consuming 1 ebit of pre-shared entanglement. After the discrimination, we still have the undisturbed second qubit, resulting in the net entanglement cost being zero, so-called {\it entanglement discrimination catalysis}.

This catalytic usage of pre-shared entanglement also can be done with our strategy. Let us consider the sign table of the set as
\begin{center}
\setlength{\tabcolsep}{3pt}
\renewcommand{\arraystretch}{1.2}
\begin{tabular}{ c|c|c|c|c } 
  & $s_{Z_1}$  & $s_{X_1}$ & $s_{Z_2}$ & $s_{X_2}$\\ 
 \hline
 $\ket{\Phi^{00}}$ & $+$ & $+$ & $+$ & $+$  \\ 
 $\ket{\Phi^{11}}$ & $+$ & $-$ & $+$ & $-$  \\ 
 $\ket{\Phi^{21}}$ & $-$ & $+$ & $+$ & $-$  \\
 $\ket{\Phi^{31}}$ & $-$ & $-$ & $+$ & $-$   \\
\end{tabular}
\end{center}
It is easy to figure out that this set can be discriminated by reading the signs of the first and second columns according to our strategy. Since the probability of each sign at each step is 1/2, the total entanglement cost is $\frac{2 \times 2 + 2 \times 2}{4} = 2$ ebits. Because our strategy is non-destructive, we end up with 2-Bell pairs (2 ebits). Thus the net entanglement cost is zero.

\subsection{Example: 6 MESs (Entanglement earning from discrimination)}

In the previous example of 4 MESs, one can achieve the perfect discrimination with consuming net zero entanglement. Interestingly, here we demonstrate a case of net positive entanglement after discrimination, so-called {\it entanglement earning from discrimination}, beyond the {\it entanglement discrimination catalysis}~\cite{yu2012four}. Suppose we have 6 MESs, a set of 3-Bell pairs such as
\begin{equation}\label{supeq:6mes}
    \left\{ \ket{\Phi^{012}}, \ket{\Phi^{021}}, \ket{\Phi^{102}}, \ket{\Phi^{120}}, \ket{\Phi^{201}}, \ket{\Phi^{210}} \right\}.
\end{equation}
The corresponding sign table is given by
\begin{center}
\setlength{\tabcolsep}{3pt}
\renewcommand{\arraystretch}{1.2}
\begin{tabular}{ c|c|c|c|c|c|c } 
  & $s_{Z_1}$ & $s_{X_1}$ & $s_{Z_2}$ & $s_{X_2}$ & $s_{Z_3}$ & $s_{X_3}$\\ 
 \hline
 $\ket{\Phi^{012}}$  & $+$ & $+$ & $+$ & $-$ & $-$ & $+$\\ 
 $\ket{\Phi^{021}}$ & $+$ & $+$ & $-$ & $+$ & $+$ & $-$\\ 
 $\ket{\Phi^{102}}$ & $+$ & $-$ & $+$ & $+$ & $-$ & $+$ \\
 $\ket{\Phi^{120}}$ & $+$ & $-$ & $-$ & $+$ & $+$ & $+$ \\
 $\ket{\Phi^{201}}$ & $-$ & $+$ & $+$ & $+$ & $+$ & $-$ \\
 $\ket{\Phi^{210}}$ & $-$ & $+$ & $+$ & $-$ & $+$ & $+$ \\
\end{tabular}
\end{center}

Let us compute the average entanglement cost of our strategy. Since all columns have the same sign balance $(1:2)$, we can choose the first column for the sign $s_{Z_1}$. If the result is `$-$', with probability $1/3$, the target state should be $\ket{\Phi^{201}}$ or $\ket{\Phi^{210}}$, so now we focus on 5th and 6th rows. Then, we can choose the 4th or 6th columns because they have balanced signs, $(1:1)$. According to the measurements for signs, we can determine the state with the probability 1/2 for each case. This costs $2$ ebits. If the result of the first column is `$+$' with probability 2/3, the target state is one of four, $\{\ket{\Phi^{012}},\ket{\Phi^{021}},\ket{\Phi^{102}},\ket{\Phi^{120}}\}$. Then we can select one among the 2nd, 3rd, or 5th columns, which have the $(1:1)$ sign balance, but others have $(1:2)$. In any case, we obtain each sign with equal probability, and it requires one more measurement to completely determine the state. This costs $3$ ebits. Consequently, the average cost is $\frac{1 \times 2 + 2 \times 3}{3}=8/3$ ebits. We can check that this cost is independent of the choice of columns with the same sign balance. Since the entanglement in the final state after the discrimination is 3 ebits, we gain 1/3 ebits on average per discrimination, namely {\it entanglement earning from discrimination}. Can we further enhance this gain? The answer is no because 8/3 ebits is the optimal entanglement cost for our strategy. The proof is provided in the next section.

Meanwhile, we demonstrate that the conventional approach to local state discrimination cannot achieve this positive net entanglement gain. 
Before computing the entanglement cost for the perfect discrimination of the set given in Eq.~(\ref{supeq:6mes}), let us consider the local distinguishability of the set. From the upper bound on the success probability of local discrimination of $k$ MESs~\cite{nathanson2005distinguishing}, i.e., $p_\text{succ}\leq d/k$ with $d=8$ and $k=6$ in our case, we cannot rule out a possibility that the set can be perfectly distinguished by LOCC. The point is that even if the local discrimination of the set is possible, we end up with no entanglement after the discrimination, resulting in no gain. 

Next, we compute the entanglement cost for discriminating the set using pre-shared entanglement. Since any Bell pair can be one of three Bell states, we should use teleportation to distinguish them, resulting in 1 ebit of cost. For example, when we perform the teleportation on the first qubit, the possible target state is $\{ \ket{\Phi^{012}},\ket{\Phi^{021}}\}$, $\{ \ket{\Phi^{102}},\ket{\Phi^{120}}\}$, or $\{ \ket{\Phi^{201}},\ket{\Phi^{210}}\}$, according to the measurement outcomes of the first Bell pair as $\ket{\Phi^0}$, $\ket{\Phi^1}$, or $\ket{\Phi^2}$, respectively. Then we can discriminate them by performing local measurements on the second or third qubit. Hence, the total entanglement cost is 1 ebit, and the remaining state is a single Bell pair. Consequently, the net entanglement cost is zero, which has a gap of 1/3 ebits compared with our strategy.

We note that the gap originates from the advantage of the stabilizer method over teleportation in discriminating a qubit among three Bell states. In other words, this gap can be removed if we do not need to perform a teleportation to discriminate any three Bell states. This might be possible because we can freely apply local unitary operations to the set, transforming it into another set of MESs with the desired property.

Suppose we have $\{ \ket{\Phi^{000}}, \ket{\Phi^{010}},\ket{\Phi^{100}},\ket{\Phi^{110}},\ket{\Phi^{200}},\ket{\Phi^{300}}\}$ via such a transformation, where the first qubit could be one of 4 Bell states. Let us perform the teleportation on the first qubit, which can distinguish 4 Bell states by consuming 1 ebit of pre-shared entanglement. Then if the outcomes of Bell measurement on the first qubit are $\ket{\Phi^2}$ or $\ket{\Phi^3}$, it directly specifies the target state as $\ket{\Phi^{200}}$ or $\ket{\Phi^{300}}$, respectively, with preserving 2 ebits of the entanglement. If the outcomes are $\ket{\Phi^0}$ or $\ket{\Phi^1}$, we need to perform a local measurement on the second qubit to determine the state, resulting in the remaining 1 ebit. Therefore, the net entanglement cost is $\frac{1 \times 1 + 2 \times 0}{3}=1/3$ ebits. Note that if any state appears three times in the first qubit, e.g., $\{ \ket{\Phi^{000}}, \ket{\Phi^{010}},\ket{\Phi^{001}},\ket{\Phi^{110}},\ket{\Phi^{200}},\ket{\Phi^{300}}\}$, We cannot achieve 1/3 ebits because these three states must be discriminated in the second and third qubits through local measurements. This results in 1 ebit with a probability of 1/3 and zero entanglement with a probability of 2/3 after the discrimination. Since 1 ebit is used for the teleportation, the net entanglement cost becomes $\frac{1\times 1}{2}+\frac{1}{2}(\frac{1}{3} \times 0+\frac{2}{3} \times (-1))=1/6 $ ebits.

We will show that the transformation of the set mentioned above cannot be achieved through local unitary operations. Note that a set of $k$ MESs can be represented by a set of $k$ unitary matrices $U_i$ acting on the local system $A$ with $i=1,\dots,k$ such that $U_{i,A} \otimes \mathbb{1}_B\ket{\Phi}_{AB}$ for a reference MES $\ket{\Phi}_{AB}$. We borrow a result from linear algebra without proof.

\begin{theorem}[\cite{jing2015unitary}]\label{th:1} 
Let $\{ U_i\}$ and $\{ V_i\}$ be two sets of unitary matrices. Then $\{ U_i\}$ and $\{ V_i\}$ are unitary similar if and only if $\Tr[w(\{U_i\})]=\Tr[w(\{V_i\})]$ for any word w.
\end{theorem}
Here, the unitary similarity means that there exist two unitary matrices $M$ and $N$ satisfying $N^{\dagger}U_i M=V_i$ for all $i$ and the word $w(S)$ for a set of matrices $S$ is a matrix product $xy\cdots z$, where $x,y,\cdots,z$ are arbitrary matrices in the set $S$. In our case,  if we set the reference state as $\ket{\Phi^{000}}$, the 6 MESs can be represented by local Pauli operators (unitary matrices) acting on the system $A$ such that
\begin{align*}
\ket{\Phi^{012}} &= I_{A_1} Z_{A_2} X_{A_3} \ket{\Phi^{000}},\\
\ket{\Phi^{021}} &= I_{A_1} X_{A_2} Z_{A_3} \ket{\Phi^{000}},\\
\ket{\Phi^{102}} &= Z_{A_1} I_{A_2} X_{A_3} \ket{\Phi^{000}},\\
\ket{\Phi^{120}} &= Z_{A_1} X_{A_2} I_{A_3} \ket{\Phi^{000}},\\
\ket{\Phi^{201}} &= X_{A_1} I_{A_2} Z_{A_3} \ket{\Phi^{000}},\\
\ket{\Phi^{210}} &= X_{A_1} Z_{A_2} I_{A_3} \ket{\Phi^{000}}.   
\end{align*}

Then, $\{ U_i\}=\{  I_{A_1} Z_{A_2} X_{A_3},  I_{A_1} X_{A_2} Z_{A_3}, Z_{A_1} I_{A_2} X_{A_3}, Z_{A_1} X_{A_2} I_{A_3}, X_{A_1} I_{A_2} Z_{A_3}, X_{A_1} Z_{A_2} I_{A_3}\}$, and our task is to check the unitary similar condition between $\{ U_i\}$ and $\{ V_i \}$, where the latter is a set of Pauli operators achieving the same net entanglement cost with ours. First, we note that the only word giving a nonzero trace is $I_{A_1}I_{A_2}I_{A_3}$ because all Pauli matrices are traceless. In our example, one can figure out that a word of the product of all elements yields this, i.e., $\Pi_{i=1}^k U_i=I_{A_1}I_{A_2}I_{A_3}$.

Suppose the first qubit can be one of 4 Bell states, whose set of Pauli operators is $\{I,X,Y,Z,X,Y\}_{A_1}$, where any state does not appear three times for the net entanglement cost of 1/3 ebits. The all-product word contains the Pauli operator $Z_{A_1}$, resulting in a zero trace. Since the traces for this all-product word are different, the two sets are unitary inequivalent according to the Theorem~\ref{th:1}. Therefore, our set of 6 MESs cannot be locally transformed into sets that achieve the same net entanglement cost of $1/3$ ebits using the conventional teleportation-based strategy. 

There are two remarks here. First, there is still a possibility for an entanglement gain of 1/6 ebits in the conventional scheme even though it cannot reach the same cost as ours. Second, our strategy is adaptive and non-destructive, but we have not specified which property provides the advantage. We conjecture that adaptiveness is more crucial, but leave it as an open question.

\subsection{Lower bound on the  entanglement cost (Proof of theorem~\ref{th:lower-bound})}

In this section, we show that the entanglement cost for the the perfect NDLD in the previous example, 8/3 ebits, is optimal for the stabilizer method. To do that, we provide a lower bound on the average entanglement cost of the stabilizer method for a given number of $n$-Bell pairs. 

We first note that for a given set of $n$-Bell pairs, one can unambiguously identify the target state by reading a sufficient number of signs in the sign table. This can be regarded as imposing a uniquely decodable code on the set, where the sign tuples corresponds to codewords with length $2n$. Since each reading of a sign costs 1 ebit of pre-shared entanglement, the average entanglement cost depends on the average codeword length for the set. Thus, we can obtain a lower bound on the entanglement cost for the set from a lower bound on the average codeword length.

Here, we choose a prefix code as a type of uniquely decodable code. A prefix code is a code where no codeword in the system is a prefix of any other codeword. For instance, for equally distributed three Bell states $\{ \ket{\Phi^0}, \ket{\Phi^1},\ket{\Phi^2} \}$, we can assign $0$ to $\ket{\Phi^2}$, $10$ to $\ket{\Phi^0}$, and $11$ to $\ket{\Phi^1}$ for a prefix code as
\begin{center}
\setlength{\tabcolsep}{3pt}
\renewcommand{\arraystretch}{1.2}
\begin{tabular}{ c|c|c|c } 
  & $s_Z$ & $s_X$ & prefix code\\ 
 \hline
 $\ket{\Phi^{0}}$  & $+$ & $+$ & 10\\ 
 $\ket{\Phi^{1}}$ & $+$ & $-$ & 11\\ 
 $\ket{\Phi^{2}}$ & $-$ & $+$ & 0 \\
\end{tabular}
\end{center}
As seen in the table, the codeword lengths $\{1, 2, 2\}$ match the number of sign readings needed to determine the target state if we first read $s_Z$. Dividing by the number of states $k$, this gives the average entanglement cost, $\frac{1+2+2}{3}=\frac{5}{3}$ ebits. Can we assign 00 to $\ket{\Phi^2}$ instead of single 0? 
This is possible because it satisfies the condition of the prefix code, but results in redundant coding. In other words, this implies reading both signs $s_Z$ and $s_X$ for $\ket{\Phi^2}$, which is not an efficient strategy because it costs $\frac{2+2+2}{3}=2$ ebits of entanglement cost, larger than $5/3$ ebits. Then, our question is: What is the most efficient strategy? Equivalently, what is the lowest average codeword length for a given set?

To answer this, let us consider the Kraft-McMillan inequality~\cite{kraft1949device,mcmillan1956two}, which gives a necessary and sufficient condition for the existence of a prefix code for a given set of codeword lengths. For a set of codeword lengths $\{l_1,\dots,\l_k \}$, this is given by  
\begin{equation}
    \sum_{i=1}^k r^{-l_i}\leq 1,
\end{equation}
where $r$ is the alphabet size, i.e., $r=2$ in our case. Using these tools, we obtain a lower bound on the average entanglement cost. Here is a proof of Thoerem~\ref{th:lower-bound}:

\begin{proof}
For given number $k$ and codeword lengths $\{l_1, \dots, l_k\}$, the entanglement cost is $\frac{\sum_i^kl_i}{k}$ ebits. Thus our task is to find the minimum of the sum $\sum_i^kl_i$ over all sets of codeword lengths satisfying the Kraft-McMillan inequality. 

First, we claim that the minimum is found on the sets of codeword lengths saturating the Kraft-McMillan inequality. To prove this, suppose we have a set of codeword lengths $\{ l_i\}$, where $l_1\leq l_2 \cdots \leq l_k$ with the minimum sum $\sum_i^kl_i$ over all sets of codeword lengths, strictly holding the Kraft-McMillan inequality as $\sum_{i=1}^k r^{-l_i}< 1$. By multiplying $r^{l_k}$ on both side, the inequality becomes $r^{l_k}-\sum_{i=1}^k r^{l_k-l_i}>0$. Since each term in LHS is a natural number, this implies $r^{l_k}-\sum_{i=1}^k r^{l_k-l_i}\geq1$. Consequently, we get $1-\sum_{i=1}^k r^{-l_i}\geq r^{-l_k}$ after dividing $r^{l_k}$ on both side. Finally, we have inequalities given by
\begin{equation*}
    1 \geq \sum_{i=1}^{k-1}r^{-l_i}+2r^{-l_k}\geq \sum_{i=1}^{k-1}r^{-l_i}+r^{-(l_k-1)},
\end{equation*}
where the second inequality is saturated when $r=2$.
This implies we have another set of codeword lengths $\{l_1,l_2,\dots,l_{k-1},l_k-1\}$ satisfying the Kraft-McMillan inequality, with a sum of codeword lengths that is strictly lower than the original set by one, leading to a contradiction.

Now, we compute the optimal codeword lengths with minimum sum $\sum_i^k l_i$ for a given $k$. Let us start a case of $k=2^m$ with a positive integer $m$. The corresponding optimal codeword lengths are all $m$'s, i.e., $\{m,\dots,m\}$, because the average codeword length reaches the Shannon entropy as $\frac{\sum_i^{k}l_i}{k}=-\sum\frac{1}{2^m}\log_2\frac{1}{2^m}=m$~\cite{tomamichel2022information}. Next, we consider a case of $k=2^m+1$. A candidate is $\{m,\dots,m,m+1,m+1 \}$, where the number of $m$'s is $2^m-1$, saturating the Kraft-McMillan inequality. To check the optimality, suppose we have a codeword with the length of $m-1$ in the set. To compensate for this codeword, we must use four codewords with the length of $m+1$ to saturate the Kraft-McMillan inequality. Then our codeword lengths become $\{m-1,m,\dots,m,m+1,m+1,m+1,m+1\}$ with $2^m-4$ of $m$'s. Compared with $\{m,\dots,m,m+1,m+1 \}$, the sum of codeword lengths is greater by 1, so thus we cannot use the codeword length of $m-1$. Similarly, we cannot use the codeword length of $m-2$ for the optimal codeword lengths, and so on. This implies the candidate $\{m,\dots,m,m+1,m+1 \}$ are the optimal codeword lengths. By the same reasoning, for $k=2^m+t$ with $t \in [1,2^m-1]$, the optimal codeword lengths are $\{m,\dots,m,m+1,\dots,m+1 \}$ where the number of $m$'s is $2^m-t$ and of $(m+1)$'s is $2t$.

Based on this result, we compute a lower bound on the entanglement cost for $k$ of $n$-Bell pairs, which is given by
\begin{equation*}
    \min_{\{l_i\}}\frac{\sum_i^kl_i}{k}=\frac{(2^m-t)m+2t(m+1)}{k}=\frac{\{2^m-(k-2^m)\}m+2(k-2^m)(m+1)}{k}=m+2-\frac{2^{m+1}}{k}~~\text{(ebits)},
\end{equation*}
where $m=\lfloor\log_2 k\rfloor$.
\end{proof}

In the previous example, for $k=6$, this lower bound becomes 8/3 ebits, proving that our strategy achieves the optimal entanglement cost.

\section{Non-destructive local discrimination of GHZ states}\label{app:GHZ}

Our strategy based on the stabilizer formalism for the perfect NDLD can be extended to multipartite scenarios. We illustrate this by investigating the pre-shared entanglement costs for the perfect NDLD of GHZ states, which can be determined by the signs of stabilizer generators $Z_1 Z_2$, $Z_2 Z_3$, and $X_1 X_2 X_3$. Similar to the bipartite case, we consider NDLD of three GHZ states with the same sign of $X_1 X_2 X_3$. Those states can be perfectly distinguished by using LOCC~\cite{hayashi2006bounds}. Therefore, in the conventional discrimination-and-repreparation strategy using local discrimination, the entanglement cost is 1 GHZ state for the repreparation, equivalent to 2 ebits. If one adopts teleportation, the entanglement cost is always equal to or higher than $2$ ebits as at least an additional $1$ ebit is required for the repreparation. In contrast, the average entanglement cost using the non-local parity measurement is $5/3$ ebits, following the same arguments of the bipartite case to determine the signs of $Z_1 Z_2$ and $Z_2 Z_3$. Hence, a strict gap exists between the two strategies even in a multipartite scenario. While this observation can be further generalized for any stabilizer states, characterizing the advantage of the non-destructive strategy becomes more complicated when determining the signs of stabilizer generators with more than two non-local parties, for example, $X_1 X_2 X_3$.

\section{Entanglement certification under depolarizing noise}\label{app:cert}

Let us consider our certification protocol when communication channels and pre-shared entanglement are contaminated by depolarizing noise. If the initial state is an MES within a known set of MESs $\{ \ket{\Psi^z}\}_{z=1}^k$, the state after being transmitted through the depolarizing channel to Alice and Bob is given by
\begin{equation}
\rho_{\eta}^z={\cal N}_{\eta}(\ket{\Psi_z}\bra{\Psi_z})=(1-\eta)\ket{\Psi_z}\bra{\Psi_z}+\frac{\eta}{d^2}\mathbb{1},
\end{equation}
where $\eta \in [0,1]$ is the noise parameter of the channel.
In Appendix B, we consider a noiseless pre-shared entangled state with the dimension $f$. Here, we assume Alice and Bob have a noisy pre-shared entanglement $\rho_{\tau}^{\otimes n}$ with
\begin{equation}
    \rho_{\tau}=(1-\tau)\ket{\Phi^0}\bra{\Phi^0}+\frac{\tau}{d^2}\mathbb{1},
\end{equation}
where $\tau \in [0,1]$ and $\ket{\Phi^0}=1/\sqrt{d}\sum_{i=0}^{d-1}\ket{ii}$. The state $\rho^z_{\eta}$ and $\rho_\tau$ can be rewritten as
\begin{align}
    \rho_{\eta}^z&=\left(1-\frac{d^2-1}{d^2}\eta\right)\ket{\Psi_z}\bra{\Psi_z}+\frac{\eta}{d^2}(\mathbb{1}-\ket{\Psi_z}\bra{\Psi_z}), \\
    \rho_{\tau}&=\left(1-\frac{d^2-1}{d^2}\tau\right)\ket{\Phi^0}\bra{\Phi^0}+\frac{\tau}{d^2}(\mathbb{1}-\ket{\Phi^0}\bra{\Phi^0}).
\end{align}
After the local discrimination using pre-shared entanglement, the referee receives the state back via the depolarizing channel again and performs a projective measurement onto the initial state, i.e., $\{ \ket{\Psi^z}\bra{\Psi^z}\}$. Then the total score of the certification protocol is given by
\begin{align}
    {\cal S}&=\frac{1}{k}\sum_{z,\lambda_z}p(\lambda|z)\Tr \left[ \ket{\Psi_z}\bra{\Psi_z}{\cal N}_{\eta}\left(K^{\lambda_z}_{AA'}\otimes K^{\lambda_z}_{BB'} \rho^z_{\eta}\otimes \rho_{\tau}^{\otimes n}K^{\lambda_z\dagger}_{AA'}\otimes K^{\lambda_z\dagger}_{BB'}/p(\lambda|z)\right)\right]\\
    &=\frac{1}{k}\sum_{z,\lambda_z}\Tr \left[ {\cal N}_{\eta}\left(\ket{\Psi_z}\bra{\Psi_z}\right)K^{\lambda_z}_{AA'}\otimes K^{\lambda_z}_{BB'} \rho^z_{\eta}\otimes \rho_{\tau}^{\otimes n}K^{\lambda_z\dagger}_{AA'}\otimes K^{\lambda_z\dagger}_{BB'}\right]\\
    &=\left(1-\frac{d^2-1}{d^2}\eta\right)^2\left(1-\frac{d^2-1}{d^2}\tau\right)^n\frac{1}{k}\sum_{z,\lambda_z}\Tr \left[ \ket{\Psi_z}\bra{\Psi_z}K^{\lambda_z}_{AA'}\otimes K^{\lambda_z}_{BB'} \ket{\Psi_z}\bra{\Psi_z}\otimes \ket{\Phi^0}\bra{\Phi^0}^{ \otimes n}K^{\lambda_z\dagger}_{AA'}\otimes K^{\lambda_z\dagger}_{BB'}\right] \nonumber\\
    &+(\text{error terms})\\
    &=\left(1-\frac{d^2-1}{d^2}\eta\right)^2\left(1-\frac{d^2-1}{d^2}\tau\right)^{ n}{\cal S}_0+{\cal S}_e,
\end{align}
where ${\cal S}_0$ is the total score of the noiseless case and ${\cal S}_e\geq 0$ is the extra score coming from the error cases. To obtain the unit score for the noiseless case ${\cal S}_0=1$, two copies of pre-shared MESs ($n=2$) are sufficient. For example, an MES is needed for the (high-dimensional) quantum teleportation~\cite{luo2019quantum} to localize $\ket{\Psi_z}$, and the other MES is used for the repreparation of the state. 

Let us examine a special case of noiseless channel, i.e., $\eta=0$. Then if we adopt the same strategy  with the noiseless case, a sufficient condition for the certification of entanglement from the condition ${\cal S}>\frac{1}{k}$ is given by
\begin{equation}
    \left(1-\frac{d^2-1}{d^2}\tau\right)^2>\frac{1}{k},
\end{equation}
resulting in $\tau<\left( 1-\frac{1}{\sqrt{k}}\right)\frac{d^2}{d^2-1}$. When $k=d^2$, this condition yields $\tau<\frac{d}{d+1}$, which implies we can certify the entanglement of any entangled Werner-like state. Note that nonlocality test-based certification protocols cannot detect all entangled Werner-like states~\cite{werner1989quantum, barret2002nonsequential, almeida2007noise, oszmaniec2017simulating, hirsch2017betterlocalhidden}. Conversely, when the pre-shared entanglement is noiseless ($\tau=0$), our protocol is valid as long as the states evolved under the channel are entangled ($\eta<\frac{d}{d+1}$). 

\subsection{Example: 2 Bell states}
We explicitly illustrate the cases of Bell states ($d=2$), by computing the achievable score including the noisy term ${\cal S}_e$ using the stabilizer method. To do that, let us express a general Pauli noise channel as
\begin{equation}
    {\cal N}(\rho)=\sum_{i_X,i_Z=0}^1q_{i_Xi_Z}X^{i_X}Z^{i_Z}\rho Z^{i_Z}X^{i_X},
\end{equation}
where $X^{i_X}$ and $Z^{i_Z}$ are local Pauli operators acting on system $B$ and $\sum_{i_Xi_Z} q_{i_Xi_Z}=1$. When the pre-shared entanglement is contaminated under the Pauli error, the effect of the corresponding non-local parity measurement can be written as
\begin{equation}
    U_{P_A} \otimes U_{P_B} \ket{\Psi_z}_{AB}  (\mathbb{1}_{A'}\otimes X_{B'}^{i_X}Z_{B'}^{i_Z})\ket{\Phi^0}_{A'B'}\xrightarrow{\Pi^{(x,y)}_{A'B'}} \mathbb{1}_A\otimes P_B^{i_X}\frac{\mathbb{1}_{AB}+(-1)^{x+y+i_Z}P_AP_B}{2} \ket{\Psi_z}_{AB}, 
\end{equation}
$\Pi_{A'B'}^{(x,y)} = H^{\otimes 2} \left( \Pi^x_{A'} \otimes \Pi^y_{B'} \right) H^{\otimes 2}$, and $\Pi^x, \Pi^y$ are the measurements on the computational basis with $x,y\in \{0,1\}$.
Suppose there are two states to be distinguished, such as $\ket{\Psi_0}=\ket{\Phi^0}$ and $\ket{\Psi_1}=\ket{\Phi^2}$ without loss of generality. In the noiseless case, we can distinguish them by measuring the parity of the stabilizer generator $Z_AZ_B$ by using the pre-shared entanglement $\ket{\Phi^0}_{A'B'}$. If the Pauli noise contaminates the states and pre-shared entanglement, however, the achievable total score is given by

\begin{align}
    {\cal S}_2&=\frac{1}{2}\sum_{z=\{0,1\}}\sum_{(x,y)}\delta_{z,x\oplus y}\Tr \left[ {\cal N}\left(\ket{\Psi_z}_{AB}\bra{\Psi_z}\right)N\Pi^{(x,y)}_{A'B'}U_{P_A}\otimes U_{P_B} {\cal N}(\ket{\Psi_z}_{AB}\bra{\Psi_z})\otimes {\cal N}(\ket{\Phi^0}_{A'B'}\bra{\Phi^0})U^{\dagger}_{P_A}\otimes U^{\dagger}_{P_B}\Pi^{(x,y)}_{A'B'}\right]\\
    &=\frac{1}{2}\sum_{z=\{0,1\}}\sum_{(x,y)}\delta_{z,x\oplus y}N\Tr \left[ \sum_{ijk}q^m_{i_Xi_Z}q^s_{j_Xj_Z}q^p_{k_Xk_Z}X^{i_X}Z^{i_Z}\ket{\Psi_z}_{AB}\bra{\Psi_z}Z^{i_Z}X^{i_X}\Pi^{(x,y)}_{A'B'} U_{P_A}\otimes U_{P_B}\right.\nonumber\\&\left.\times (X^{j_X}Z^{j_Z}\ket{\Psi_z}_{AB}\bra{\Psi_z}Z^{j_Z}X^{j_X})\otimes (X^{k_X}Z^{k_Z}\ket{\Phi^0}_{A'B'}\bra{\Phi^0}Z^{k_Z}X^{k_X})
     U^{\dagger}_{P_A}\otimes U^{\dagger}_{P_B}\Pi^{(x,y)}_{A'B'}\right]\\
    &=\frac{1}{2}\sum_{z=\{0,1\}}\sum_{(x,y)}\delta_{z,x\oplus y}\Tr \left[ \sum_{ijk}q^m_{i_Xi_Z}q^s_{j_Xj_Z}q^p_{k_Xk_Z}X^{i_X}Z^{i_Z}\ket{\Psi_z}_{AB}\bra{\Psi_z}Z^{i_Z}X^{i_X} Z_B^{k_X}\frac{\mathbb{1}_{AB}+(-1)^{x+y+k_Z}Z_AZ_B}{2} \right.\nonumber\\
    &\left.\times X^{j_X}Z^{j_Z}\ket{\Psi_z}_{AB}\bra{\Psi_z}Z^{j_Z}X^{j_X}\frac{\mathbb{1}_{AB}+(-1)^{x+y+k_Z}Z_AZ_B}{2} Z_B^{k_X}\right]\\
    &=\frac{1}{2}\sum_{z=\{0,1\}}\sum_{(x,y)}\delta_{z,x\oplus y}\Tr \left[ \sum_{ijk}q^m_{i_Xi_Z}q^s_{j_Xj_Z}q^p_{k_Xk_Z}X^{i_X}Z^{i_Z}\ket{\Psi_z}_{AB}\bra{\Psi_z}Z^{i_Z}X^{i_X} Z^{k_X}\left( \frac{1+e^{i\pi(x+y+z+k_Z+j_X)}}{2}\right) \right.\nonumber\\
    &\left.\times X^{j_X}Z^{j_Z}\ket{\Psi_z}_{AB}\bra{\Psi_z}Z^{j_Z}X^{j_X}Z^{k_X}\right]\\
    &=\sum_{z=\{0,1\}}\sum_{ijk}q^m_{i_Xi_Z}q^s_{j_Xj_Z}q^p_{k_Xk_Z}\left( \frac{1+e^{i\pi(z+k_Z+j_X)}}{2}\right) \Tr \left[ Z^{j_Z}X^{j_X}Z^{k_X}X^{i_X}Z^{i_Z}\ket{\Psi_z}_{AB}\bra{\Psi_z}Z^{i_Z}X^{i_X}Z^{k_X}X^{j_X}Z^{j_Z} \nonumber \right.\\
    &\left. \times \ket{\Psi_z}_{AB}\bra{\Psi_z}\right],
\end{align}
where $N$ is an appropriate normalization constant, $(q^m_{i_Xi_Z}, X^{i_X}Z^{i_Z})$, $(q^s_{j_Xj_Z}, X^{j_X}Z^{j_Z})$, and $(q^m_{k_Xk_Z}, X^{k_X}Z^{k_Z})$ represent the channel noises for the measurement, input state, and pre-shared entanglement, respectively, and $\sum_{ijk}$ is the summation over all $ i_X,i_Z,j_X,j_Z,k_X,k_Z \in \{ 0,1\}$ and $ \oplus $ computed under modulus 2. The condition for the nonvanishing terms in the summation is given by
\begin{align}
     k_Z\oplus j_X =0~\wedge~ i_X\oplus j_X=0~\wedge~ i_Z\oplus k_X\oplus j_Z=0,
\end{align}
where the trace term is 1 when the condition is satisfied. For a depolarizing channel, the corresponding coefficients are $q^m_{00}=q^s_{00}=1-\frac{3}{4}\eta$, $q_{00}^p=1-\frac{3}{4}\tau$, and $q^m_{i_Xi_Z}=q^s_{j_Xj_Z}=\frac{1}{4}\eta$, $q^p_{k_Xk_Z}=\frac{1}{4}\tau$ when $(i_Xi_Z),(j_Xj_Z),(k_Xk_Z)\neq (00)$. Consequently, the total score is given by
\begin{equation}
    {\cal S}_2=\left(1-\frac{3}{4}\eta\right)^2\left(1-\frac{3}{4}\tau\right)+\left(\frac{\eta}{4}\right)^2\left(1-\frac{3}{4}\tau\right)+2\left(1-\frac{3}{4}\eta\right)\frac{\eta}{4}\frac{\tau}{4}+4\left(\frac{\eta}{4}\right)^2\frac{\tau}{4}.
\end{equation}
For the noiseless channel ($\eta=0$), this achievable score is larger than the local bound of $1/2$ when $\tau<2/3$, which is satisfied as long as the pre-shared Werner-like state is entangled. On the other hand, even if we have a noiseless pre-shared entanglement, we need a slightly lower noise channel with $\eta<2/5$ to surpass the local bound.

\subsection{Example: 4 Bell states}

Next, we examine the $k=4$ case involving all four Bell states such that $\ket{\Psi_{z_Xz_Z}}=X^{z_X}Z^{z_Z}\ket{\Phi^{0}}$ with $z_X, z_Z \in \{0,1\}$. Then the achievable score using the stabilizer method is given by 
\begin{align}
    {\cal S}_4&=\frac{1}{4}\sum_{z_X,z_Z}\sum_{\substack{(x,y)\\ (x',y')}}\delta_{z_X,x\oplus y}\delta_{z_Z,x'\oplus y'}\Tr \left[ {\cal N}\left(\ket{\Psi_z}_{AB}\bra{\Psi_z}\right)N'\Pi^{(x,y)}_{A'B'}\Pi^{(x',y')}_{A'B'}U_{P_A}\otimes U_{P_B} {\cal N}(\ket{\Psi_z}_{AB}\bra{\Psi_z})\otimes{\cal N}^{\otimes2}(\ket{\Phi^0}_{A'B'}\bra{\Phi^0}^{\otimes 2}) \right.\nonumber \\
    &\left. \times U^{\dagger}_{P_A}\otimes U^{\dagger}_{P_B}\Pi^{(x',y')}_{A'B'}\Pi^{(x,y)}_{A'B'}\right]\\
    &=\frac{1}{4}\sum_{z_X,z_Z}\sum_{\substack{(x,y)\\ (x',y')}}\delta_{z_X,x\oplus y}\delta_{z_Z,x'\oplus y'}\Tr \left[ \sum_{ijkl}q^m_{i_Xi_Z}q^s_{j_Xj_Z}q^{p_1}_{k_Xk_Z}q^{p_2}_{l_Xl_Z}X^{i_X}Z^{i_Z}\ket{\Psi_z}_{AB}\bra{\Psi_z}Z^{i_Z}X^{i_X} Z_B^{k_X}\frac{\mathbb{1}_{AB}+(-1)^{x+y+k_Z}Z_AZ_B}{2}\right. \nonumber \\
    & \left. \times X_B^{l_X}\frac{\mathbb{1}_{AB}+(-1)^{x'+y'+l_Z}X_AX_B}{2} X^{j_X}Z^{j_Z}\ket{\Psi_z}_{AB}\bra{\Psi_z}Z^{j_Z}X^{j_X}\frac{\mathbb{1}_{AB}+(-1)^{x'+y'+l_Z}X_AX_B}{2}X_B^{l_X} \right.\nonumber \\
    &\left. \times \frac{\mathbb{1}_{AB}+(-1)^{x+y+k_Z}Z_AZ_B}{2}Z_B^{k_X}\right] \\
    &=\frac{1}{4}\sum_{z_X,z_Z}\sum_{\substack{(x,y)\\ (x',y')}}\delta_{z_X,x\oplus y}\delta_{z_Z,x'\oplus y'}\Tr \left[ \sum_{ijkl}q^m_{i_Xi_Z}q^s_{j_Xj_Z}q^{p_1}_{k_Xk_Z}q^{p_2}_{l_Xl_Z}X^{i_X}Z^{i_Z}\ket{\Psi_z}_{AB}\bra{\Psi_z}Z^{i_Z}X^{i_X} Z^{k_X}\frac{1+e^{i\pi(x+y+z_X+ l_X+k_Z+j_X)}}{2}\right. \nonumber \\
    & \left. \times X^{l_X}\frac{1+e^{i\pi(x'+y'+z_Z+l_Z+j_Z)}}{2} X^{j_X}Z^{j_Z}\ket{\Psi_z}_{AB}\bra{\Psi_z}Z^{j_Z}X^{j_X}X^{l_X} Z^{k_X}\right] \\
&=\sum_{z_X,z_Z}\sum_{ijkl}q^m_{i_Xi_Z}q^s_{j_Xj_Z}q^{p_1}_{k_Xk_Z}q^{p_2}_{l_Xl_Z}\frac{1+e^{i\pi( l_X+k_Z+j_X)}}{2}\frac{1+e^{i\pi(l_Z+j_Z)}}{2}\Tr \left[ Z^{j_Z}X^{j_X}X^{l_X} Z^{k_X}X^{i_X}Z^{i_Z}\ket{\Psi_z}_{AB}\bra{\Psi_z}\right. \nonumber \\
    & \left. \times Z^{i_Z}X^{i_X} Z^{k_X} X^{l_X} X^{j_X}Z^{j_Z}\ket{\Psi_z}_{AB}\bra{\Psi_z}\right].
\end{align}
The condition for nonvanishing terms can be written as
\begin{equation}
    l_X\oplus k_Z\oplus j_X=0~\wedge~  l_Z\oplus j_Z=0 ~\wedge~i_Z \oplus k_X \oplus j_Z=0~\wedge~i_X \oplus l_X \oplus j_X=0.
\end{equation}
With the coefficients $q^m_{00}=q^s_{00}=1-\frac{3}{4}\eta$, $q_{00}^{p_1}=q_{00}^{p_2}=1-\frac{3}{4}\tau$, and $q^m_{i_Xi_Z}=q^s_{j_Xj_Z}=\frac{1}{4}\eta$ when $(i_Xi_Z), (j_Xj_Z)\neq (00)$, $q^{p_1}_{k_Xk_Z}=q^{p_2}_{l_Xl_Z}=\frac{1}{4}\tau$ when $(k_Xk_Z),(l_Xl_Z) \neq (00)$, the total score can be expressed as
\begin{align}
    {\cal S}_4&=\left( 1-\frac{3}{4}\eta\right)^2\left( 1-\frac{3}{4}\tau\right)^2+4\left(\frac{\eta}{4}\right)^2\left( 1-\frac{3}{4}\tau\right)\frac{\tau}{4}+2\left( 1-\frac{3}{4}\eta\right)\frac{\eta}{4}\left( 1-\frac{3}{4}\tau\right)\frac{\tau}{4}+4\left( 1-\frac{3}{4}\eta\right)\frac{\eta}{4}\left(\frac{\tau}{4}\right)^2+5\left(\frac{\eta}{4}\right)^2\left(\frac{\tau}{4}\right)^2.
\end{align}
Similar to the case of two Bell states, we can certify any entangled Werner-like state using a noiseless channel. Furthermore, the converse holds in this case: if the pre-shared Werner-like state is noiseless, we get a higher score than the local bound of $1/4$ as long as the states evolved under the channel remain entangled, i.e., $\eta<2/3$.

\section{Local bound on the total score for randomly generated stabilizer states}\label{app:random}
In the main text, we consider the absolutely maximally entangled states, where any bipartition results in a maximally entangled state. Note that randomly generated pure states can have such a property approximately. In specific, %the average entropy of the reduced state $\sigma=\Tr_{N/2}\ket{\psi}\bra{\psi}$ for a randomly generated pure $N$-qubit state $\ket{\psi}$ (assuming even $N$) is given by~\cite{page1993average,goyeneche2015absolutely}
%\begin{equation}
%    S(\sigma)=\frac{N}{2}-c,
%\end{equation}
for a large $N$ (assuming even $N$), the reduced state converges to a maximally mixed state. Let us consider a randomly chosen stabilizer state $\ket{\psi_S}$. Since any stabilizer state has a flat spectrum, i.e., all the Schmidt coefficients are equal, the largest Schmidt coefficient of the state is typically not far from $1/\sqrt{2^{N/2}}$ when $N\gg 1$. Especially, the probability of the entanglement of $\ket{\psi_S}$ deviating from the maximum value is given by~\cite{smith2006typical}
\begin{equation}\label{eq:random}
P\left(S(\sigma_S)<\frac{N(1-\epsilon)}{2}\right)\leq \exp\left[ -\frac{N\epsilon^2}{1024}\right], 
\end{equation}
where $\sigma_S=\Tr_{N/2}\ket{\psi_S}\bra{\psi_S}$.
Consequently, the largest eigenvalue of $\sigma_S$ is given by $\xi^2=\frac{1+\frac{N}{2}\epsilon+O(\epsilon^2)}{2^{N/2}}$ with an exponentially small failure probability as long as $\epsilon=O(\frac{1}{\sqrt{N}})$.   
Meanwhile, the total score for the $k$ randomly generated stabilizer states is bounded above by Theorem 1 as
\begin{align}
    {\cal S} &\leq \frac{1}{k} 2^N\frac{(1+\frac{N}{2}\epsilon+O(\epsilon^2))^2}{2^N}=\frac{(1+\frac{N}{2}\epsilon)^2+O(\epsilon^3)}{k}. \\
\end{align}
Therefore, if we set $\epsilon=O(\frac{1}{N})$, the bound slightly deviates from that of random guessing, i.e., $1/k$, when $N \gg 1$.

\bibliography{reference}

\begin{thebibliography}{77}%
\makeatletter
\providecommand \@ifxundefined [1]{%
 \@ifx{#1\undefined}
}%
\providecommand \@ifnum [1]{%
 \ifnum #1\expandafter \@firstoftwo
 \else \expandafter \@secondoftwo
 \fi
}%
\providecommand \@ifx [1]{%
 \ifx #1\expandafter \@firstoftwo
 \else \expandafter \@secondoftwo
 \fi
}%
\providecommand \natexlab [1]{#1}%
\providecommand \enquote  [1]{``#1''}%
\providecommand \bibnamefont  [1]{#1}%
\providecommand \bibfnamefont [1]{#1}%
\providecommand \citenamefont [1]{#1}%
\providecommand \href@noop [0]{\@secondoftwo}%
\providecommand \href [0]{\begingroup \@sanitize@url \@href}%
\providecommand \@href[1]{\@@startlink{#1}\@@href}%
\providecommand \@@href[1]{\endgroup#1\@@endlink}%
\providecommand \@sanitize@url [0]{\catcode `\\12\catcode `\$12\catcode `\&12\catcode `\#12\catcode `\^12\catcode `\_12\catcode `\%12\relax}%
\providecommand \@@startlink[1]{}%
\providecommand \@@endlink[0]{}%
\providecommand \url  [0]{\begingroup\@sanitize@url \@url }%
\providecommand \@url [1]{\endgroup\@href {#1}{\urlprefix }}%
\providecommand \urlprefix  [0]{URL }%
\providecommand \Eprint [0]{\href }%
\providecommand \doibase [0]{https://doi.org/}%
\providecommand \selectlanguage [0]{\@gobble}%
\providecommand \bibinfo  [0]{\@secondoftwo}%
\providecommand \bibfield  [0]{\@secondoftwo}%
\providecommand \translation [1]{[#1]}%
\providecommand \BibitemOpen [0]{}%
\providecommand \bibitemStop [0]{}%
\providecommand \bibitemNoStop [0]{.\EOS\space}%
\providecommand \EOS [0]{\spacefactor3000\relax}%
\providecommand \BibitemShut  [1]{\csname bibitem#1\endcsname}%
\let\auto@bib@innerbib\@empty
%</preamble>
\bibitem [{\citenamefont {Helstrom}(1969)}]{helstrom1969quantum}%
  \BibitemOpen
  \bibfield  {author} {\bibinfo {author} {\bibfnamefont {C.~W.}\ \bibnamefont {Helstrom}},\ }\bibfield  {title} {\bibinfo {title} {Quantum detection and estimation theory},\ }\href@noop {} {\bibfield  {journal} {\bibinfo  {journal} {Journal of Statistical Physics}\ }\textbf {\bibinfo {volume} {1}},\ \bibinfo {pages} {231} (\bibinfo {year} {1969})}\BibitemShut {NoStop}%
\bibitem [{\citenamefont {Holevo}(1974)}]{holevo1974remarks}%
  \BibitemOpen
  \bibfield  {author} {\bibinfo {author} {\bibfnamefont {A.~S.}\ \bibnamefont {Holevo}},\ }\bibfield  {title} {\bibinfo {title} {Remarks on optimal quantum measurements},\ }\href@noop {} {\bibfield  {journal} {\bibinfo  {journal} {Problemy Peredachi Informatsii}\ }\textbf {\bibinfo {volume} {10}},\ \bibinfo {pages} {51} (\bibinfo {year} {1974})}\BibitemShut {NoStop}%
\bibitem [{\citenamefont {Barnett}\ and\ \citenamefont {Croke}(2009)}]{barnett2009quantum}%
  \BibitemOpen
  \bibfield  {author} {\bibinfo {author} {\bibfnamefont {S.~M.}\ \bibnamefont {Barnett}}\ and\ \bibinfo {author} {\bibfnamefont {S.}~\bibnamefont {Croke}},\ }\bibfield  {title} {\bibinfo {title} {Quantum state discrimination},\ }\href@noop {} {\bibfield  {journal} {\bibinfo  {journal} {Advances in Optics and Photonics}\ }\textbf {\bibinfo {volume} {1}},\ \bibinfo {pages} {238} (\bibinfo {year} {2009})}\BibitemShut {NoStop}%
\bibitem [{\citenamefont {Bae}\ and\ \citenamefont {Kwek}(2015)}]{bae2015quantum}%
  \BibitemOpen
  \bibfield  {author} {\bibinfo {author} {\bibfnamefont {J.}~\bibnamefont {Bae}}\ and\ \bibinfo {author} {\bibfnamefont {L.-C.}\ \bibnamefont {Kwek}},\ }\bibfield  {title} {\bibinfo {title} {Quantum state discrimination and its applications},\ }\href@noop {} {\bibfield  {journal} {\bibinfo  {journal} {Journal of Physics A: Mathematical and Theoretical}\ }\textbf {\bibinfo {volume} {48}},\ \bibinfo {pages} {083001} (\bibinfo {year} {2015})}\BibitemShut {NoStop}%
\bibitem [{\citenamefont {Wootters}\ and\ \citenamefont {Zurek}(1982)}]{wootters1982single}%
  \BibitemOpen
  \bibfield  {author} {\bibinfo {author} {\bibfnamefont {W.~K.}\ \bibnamefont {Wootters}}\ and\ \bibinfo {author} {\bibfnamefont {W.~H.}\ \bibnamefont {Zurek}},\ }\bibfield  {title} {\bibinfo {title} {A single quantum cannot be cloned},\ }\href@noop {} {\bibfield  {journal} {\bibinfo  {journal} {Nature}\ }\textbf {\bibinfo {volume} {299}},\ \bibinfo {pages} {802} (\bibinfo {year} {1982})}\BibitemShut {NoStop}%
\bibitem [{\citenamefont {Bennett}\ \emph {et~al.}(1999)\citenamefont {Bennett}, \citenamefont {DiVincenzo}, \citenamefont {Fuchs}, \citenamefont {Mor}, \citenamefont {Rains}, \citenamefont {Shor}, \citenamefont {Smolin},\ and\ \citenamefont {Wootters}}]{bennett1999quantum}%
  \BibitemOpen
  \bibfield  {author} {\bibinfo {author} {\bibfnamefont {C.~H.}\ \bibnamefont {Bennett}}, \bibinfo {author} {\bibfnamefont {D.~P.}\ \bibnamefont {DiVincenzo}}, \bibinfo {author} {\bibfnamefont {C.~A.}\ \bibnamefont {Fuchs}}, \bibinfo {author} {\bibfnamefont {T.}~\bibnamefont {Mor}}, \bibinfo {author} {\bibfnamefont {E.}~\bibnamefont {Rains}}, \bibinfo {author} {\bibfnamefont {P.~W.}\ \bibnamefont {Shor}}, \bibinfo {author} {\bibfnamefont {J.~A.}\ \bibnamefont {Smolin}},\ and\ \bibinfo {author} {\bibfnamefont {W.~K.}\ \bibnamefont {Wootters}},\ }\bibfield  {title} {\bibinfo {title} {Quantum nonlocality without entanglement},\ }\href@noop {} {\bibfield  {journal} {\bibinfo  {journal} {Physical Review A}\ }\textbf {\bibinfo {volume} {59}},\ \bibinfo {pages} {1070} (\bibinfo {year} {1999})}\BibitemShut {NoStop}%
\bibitem [{\citenamefont {Schmid}\ and\ \citenamefont {Spekkens}(2018)}]{schmid2018contextual}%
  \BibitemOpen
  \bibfield  {author} {\bibinfo {author} {\bibfnamefont {D.}~\bibnamefont {Schmid}}\ and\ \bibinfo {author} {\bibfnamefont {R.~W.}\ \bibnamefont {Spekkens}},\ }\bibfield  {title} {\bibinfo {title} {Contextual advantage for state discrimination},\ }\href@noop {} {\bibfield  {journal} {\bibinfo  {journal} {Physical Review X}\ }\textbf {\bibinfo {volume} {8}},\ \bibinfo {pages} {011015} (\bibinfo {year} {2018})}\BibitemShut {NoStop}%
\bibitem [{\citenamefont {Ghosh}\ \emph {et~al.}(2001)\citenamefont {Ghosh}, \citenamefont {Kar}, \citenamefont {Roy}, \citenamefont {Sen(De)},\ and\ \citenamefont {Sen}}]{ghosh2001distinguishability}%
  \BibitemOpen
  \bibfield  {author} {\bibinfo {author} {\bibfnamefont {S.}~\bibnamefont {Ghosh}}, \bibinfo {author} {\bibfnamefont {G.}~\bibnamefont {Kar}}, \bibinfo {author} {\bibfnamefont {A.}~\bibnamefont {Roy}}, \bibinfo {author} {\bibfnamefont {A.}~\bibnamefont {Sen(De)}},\ and\ \bibinfo {author} {\bibfnamefont {U.}~\bibnamefont {Sen}},\ }\bibfield  {title} {\bibinfo {title} {Distinguishability of bell states},\ }\href@noop {} {\bibfield  {journal} {\bibinfo  {journal} {Physical Review Letters}\ }\textbf {\bibinfo {volume} {87}},\ \bibinfo {pages} {277902} (\bibinfo {year} {2001})}\BibitemShut {NoStop}%
\bibitem [{\citenamefont {Wu}\ \emph {et~al.}(2021)\citenamefont {Wu}, \citenamefont {Kondra}, \citenamefont {Rana}, \citenamefont {Scandolo}, \citenamefont {Xiang}, \citenamefont {Li}, \citenamefont {Guo},\ and\ \citenamefont {Streltsov}}]{wu2021operational}%
  \BibitemOpen
  \bibfield  {author} {\bibinfo {author} {\bibfnamefont {K.-D.}\ \bibnamefont {Wu}}, \bibinfo {author} {\bibfnamefont {T.~V.}\ \bibnamefont {Kondra}}, \bibinfo {author} {\bibfnamefont {S.}~\bibnamefont {Rana}}, \bibinfo {author} {\bibfnamefont {C.~M.}\ \bibnamefont {Scandolo}}, \bibinfo {author} {\bibfnamefont {G.-Y.}\ \bibnamefont {Xiang}}, \bibinfo {author} {\bibfnamefont {C.-F.}\ \bibnamefont {Li}}, \bibinfo {author} {\bibfnamefont {G.-C.}\ \bibnamefont {Guo}},\ and\ \bibinfo {author} {\bibfnamefont {A.}~\bibnamefont {Streltsov}},\ }\bibfield  {title} {\bibinfo {title} {Operational resource theory of imaginarity},\ }\href@noop {} {\bibfield  {journal} {\bibinfo  {journal} {Physical Review Letters}\ }\textbf {\bibinfo {volume} {126}},\ \bibinfo {pages} {090401} (\bibinfo {year} {2021})}\BibitemShut {NoStop}%
\bibitem [{\citenamefont {Terhal}\ \emph {et~al.}(2001)\citenamefont {Terhal}, \citenamefont {DiVincenzo},\ and\ \citenamefont {Leung}}]{terhal2001hiding}%
  \BibitemOpen
  \bibfield  {author} {\bibinfo {author} {\bibfnamefont {B.~M.}\ \bibnamefont {Terhal}}, \bibinfo {author} {\bibfnamefont {D.~P.}\ \bibnamefont {DiVincenzo}},\ and\ \bibinfo {author} {\bibfnamefont {D.~W.}\ \bibnamefont {Leung}},\ }\bibfield  {title} {\bibinfo {title} {Hiding bits in bell states},\ }\href@noop {} {\bibfield  {journal} {\bibinfo  {journal} {Physical Review Letters}\ }\textbf {\bibinfo {volume} {86}},\ \bibinfo {pages} {5807} (\bibinfo {year} {2001})}\BibitemShut {NoStop}%
\bibitem [{\citenamefont {DiVincenzo}\ \emph {et~al.}(2002)\citenamefont {DiVincenzo}, \citenamefont {Leung},\ and\ \citenamefont {Terhal}}]{divincenzo2002quantum}%
  \BibitemOpen
  \bibfield  {author} {\bibinfo {author} {\bibfnamefont {D.~P.}\ \bibnamefont {DiVincenzo}}, \bibinfo {author} {\bibfnamefont {D.~W.}\ \bibnamefont {Leung}},\ and\ \bibinfo {author} {\bibfnamefont {B.~M.}\ \bibnamefont {Terhal}},\ }\bibfield  {title} {\bibinfo {title} {Quantum data hiding},\ }\href@noop {} {\bibfield  {journal} {\bibinfo  {journal} {IEEE Transactions on Information Theory}\ }\textbf {\bibinfo {volume} {48}},\ \bibinfo {pages} {580} (\bibinfo {year} {2002})}\BibitemShut {NoStop}%
\bibitem [{\citenamefont {Eggeling}\ and\ \citenamefont {Werner}(2002)}]{eggeling2002hiding}%
  \BibitemOpen
  \bibfield  {author} {\bibinfo {author} {\bibfnamefont {T.}~\bibnamefont {Eggeling}}\ and\ \bibinfo {author} {\bibfnamefont {R.~F.}\ \bibnamefont {Werner}},\ }\bibfield  {title} {\bibinfo {title} {Hiding classical data in multipartite quantum states},\ }\href@noop {} {\bibfield  {journal} {\bibinfo  {journal} {Physical Review Letters}\ }\textbf {\bibinfo {volume} {89}},\ \bibinfo {pages} {097905} (\bibinfo {year} {2002})}\BibitemShut {NoStop}%
\bibitem [{\citenamefont {Matthews}\ \emph {et~al.}(2009)\citenamefont {Matthews}, \citenamefont {Wehner},\ and\ \citenamefont {Winter}}]{matthews2009distinguishability}%
  \BibitemOpen
  \bibfield  {author} {\bibinfo {author} {\bibfnamefont {W.}~\bibnamefont {Matthews}}, \bibinfo {author} {\bibfnamefont {S.}~\bibnamefont {Wehner}},\ and\ \bibinfo {author} {\bibfnamefont {A.}~\bibnamefont {Winter}},\ }\bibfield  {title} {\bibinfo {title} {Distinguishability of quantum states under restricted families of measurements with an application to quantum data hiding},\ }\href@noop {} {\bibfield  {journal} {\bibinfo  {journal} {Communications in Mathematical Physics}\ }\textbf {\bibinfo {volume} {291}},\ \bibinfo {pages} {813} (\bibinfo {year} {2009})}\BibitemShut {NoStop}%
\bibitem [{\citenamefont {Markham}\ and\ \citenamefont {Sanders}(2008)}]{markham2008graph}%
  \BibitemOpen
  \bibfield  {author} {\bibinfo {author} {\bibfnamefont {D.}~\bibnamefont {Markham}}\ and\ \bibinfo {author} {\bibfnamefont {B.~C.}\ \bibnamefont {Sanders}},\ }\bibfield  {title} {\bibinfo {title} {Graph states for quantum secret sharing},\ }\href@noop {} {\bibfield  {journal} {\bibinfo  {journal} {Physical Review A}\ }\textbf {\bibinfo {volume} {78}},\ \bibinfo {pages} {042309} (\bibinfo {year} {2008})}\BibitemShut {NoStop}%
\bibitem [{\citenamefont {Rahaman}\ and\ \citenamefont {Parker}(2015)}]{rahaman2015quantum}%
  \BibitemOpen
  \bibfield  {author} {\bibinfo {author} {\bibfnamefont {R.}~\bibnamefont {Rahaman}}\ and\ \bibinfo {author} {\bibfnamefont {M.~G.}\ \bibnamefont {Parker}},\ }\bibfield  {title} {\bibinfo {title} {Quantum scheme for secret sharing based on local distinguishability},\ }\href@noop {} {\bibfield  {journal} {\bibinfo  {journal} {Physical Review A}\ }\textbf {\bibinfo {volume} {91}},\ \bibinfo {pages} {022330} (\bibinfo {year} {2015})}\BibitemShut {NoStop}%
\bibitem [{\citenamefont {Banik}\ \emph {et~al.}(2021)\citenamefont {Banik}, \citenamefont {Guha}, \citenamefont {Alimuddin}, \citenamefont {Kar}, \citenamefont {Halder},\ and\ \citenamefont {Bhattacharya}}]{banik2021multicopy}%
  \BibitemOpen
  \bibfield  {author} {\bibinfo {author} {\bibfnamefont {M.}~\bibnamefont {Banik}}, \bibinfo {author} {\bibfnamefont {T.}~\bibnamefont {Guha}}, \bibinfo {author} {\bibfnamefont {M.}~\bibnamefont {Alimuddin}}, \bibinfo {author} {\bibfnamefont {G.}~\bibnamefont {Kar}}, \bibinfo {author} {\bibfnamefont {S.}~\bibnamefont {Halder}},\ and\ \bibinfo {author} {\bibfnamefont {S.~S.}\ \bibnamefont {Bhattacharya}},\ }\bibfield  {title} {\bibinfo {title} {Multicopy adaptive local discrimination: Strongest possible two-qubit nonlocal bases},\ }\href@noop {} {\bibfield  {journal} {\bibinfo  {journal} {Physical Review Letters}\ }\textbf {\bibinfo {volume} {126}},\ \bibinfo {pages} {210505} (\bibinfo {year} {2021})}\BibitemShut {NoStop}%
\bibitem [{\citenamefont {Kribs}\ \emph {et~al.}(2019)\citenamefont {Kribs}, \citenamefont {Mintah}, \citenamefont {Nathanson},\ and\ \citenamefont {Pereira}}]{kribs2019quantum}%
  \BibitemOpen
  \bibfield  {author} {\bibinfo {author} {\bibfnamefont {D.~W.}\ \bibnamefont {Kribs}}, \bibinfo {author} {\bibfnamefont {C.}~\bibnamefont {Mintah}}, \bibinfo {author} {\bibfnamefont {M.}~\bibnamefont {Nathanson}},\ and\ \bibinfo {author} {\bibfnamefont {R.}~\bibnamefont {Pereira}},\ }\bibfield  {title} {\bibinfo {title} {Quantum error correction and one-way locc state distinguishability},\ }\href@noop {} {\bibfield  {journal} {\bibinfo  {journal} {Journal of Mathematical Physics}\ }\textbf {\bibinfo {volume} {60}} (\bibinfo {year} {2019})}\BibitemShut {NoStop}%
\bibitem [{\citenamefont {Walgate}\ \emph {et~al.}(2000)\citenamefont {Walgate}, \citenamefont {Short}, \citenamefont {Hardy},\ and\ \citenamefont {Vedral}}]{walgate2000local}%
  \BibitemOpen
  \bibfield  {author} {\bibinfo {author} {\bibfnamefont {J.}~\bibnamefont {Walgate}}, \bibinfo {author} {\bibfnamefont {A.~J.}\ \bibnamefont {Short}}, \bibinfo {author} {\bibfnamefont {L.}~\bibnamefont {Hardy}},\ and\ \bibinfo {author} {\bibfnamefont {V.}~\bibnamefont {Vedral}},\ }\bibfield  {title} {\bibinfo {title} {Local distinguishability of multipartite orthogonal quantum states},\ }\href@noop {} {\bibfield  {journal} {\bibinfo  {journal} {Physical Review Letters}\ }\textbf {\bibinfo {volume} {85}},\ \bibinfo {pages} {4972} (\bibinfo {year} {2000})}\BibitemShut {NoStop}%
\bibitem [{\citenamefont {Nathanson}(2005)}]{nathanson2005distinguishing}%
  \BibitemOpen
  \bibfield  {author} {\bibinfo {author} {\bibfnamefont {M.}~\bibnamefont {Nathanson}},\ }\bibfield  {title} {\bibinfo {title} {Distinguishing bipartitite orthogonal states using locc: Best and worst cases},\ }\href@noop {} {\bibfield  {journal} {\bibinfo  {journal} {Journal of Mathematical Physics}\ }\textbf {\bibinfo {volume} {46}},\ \bibinfo {pages} {062103} (\bibinfo {year} {2005})}\BibitemShut {NoStop}%
\bibitem [{\citenamefont {Yu}\ \emph {et~al.}(2012)\citenamefont {Yu}, \citenamefont {Duan},\ and\ \citenamefont {Ying}}]{yu2012four}%
  \BibitemOpen
  \bibfield  {author} {\bibinfo {author} {\bibfnamefont {N.}~\bibnamefont {Yu}}, \bibinfo {author} {\bibfnamefont {R.}~\bibnamefont {Duan}},\ and\ \bibinfo {author} {\bibfnamefont {M.}~\bibnamefont {Ying}},\ }\bibfield  {title} {\bibinfo {title} {Four locally indistinguishable ququad-ququad orthogonal maximally entangled states},\ }\href@noop {} {\bibfield  {journal} {\bibinfo  {journal} {Physical Review Letters}\ }\textbf {\bibinfo {volume} {109}},\ \bibinfo {pages} {020506} (\bibinfo {year} {2012})}\BibitemShut {NoStop}%
\bibitem [{\citenamefont {Nathanson}(2013)}]{nathanson2013three}%
  \BibitemOpen
  \bibfield  {author} {\bibinfo {author} {\bibfnamefont {M.}~\bibnamefont {Nathanson}},\ }\bibfield  {title} {\bibinfo {title} {Three maximally entangled states can require two-way local operations and classical communication for local discrimination},\ }\href@noop {} {\bibfield  {journal} {\bibinfo  {journal} {Physical Review A}\ }\textbf {\bibinfo {volume} {88}},\ \bibinfo {pages} {062316} (\bibinfo {year} {2013})}\BibitemShut {NoStop}%
\bibitem [{\citenamefont {Cosentino}\ and\ \citenamefont {Russo}(2014)}]{10.5555/2685164.2685167}%
  \BibitemOpen
  \bibfield  {author} {\bibinfo {author} {\bibfnamefont {A.}~\bibnamefont {Cosentino}}\ and\ \bibinfo {author} {\bibfnamefont {V.}~\bibnamefont {Russo}},\ }\bibfield  {title} {\bibinfo {title} {Small sets of locally indistinguishable orthogonal maximally entangled states},\ }\href@noop {} {\bibfield  {journal} {\bibinfo  {journal} {Quantum Infomation and Computation}\ }\textbf {\bibinfo {volume} {14}},\ \bibinfo {pages} {1098–1106} (\bibinfo {year} {2014})}\BibitemShut {NoStop}%
\bibitem [{\citenamefont {Li}\ \emph {et~al.}(2015)\citenamefont {Li}, \citenamefont {Wang}, \citenamefont {Fei},\ and\ \citenamefont {Zheng}}]{li2015d}%
  \BibitemOpen
  \bibfield  {author} {\bibinfo {author} {\bibfnamefont {M.-S.}\ \bibnamefont {Li}}, \bibinfo {author} {\bibfnamefont {Y.-L.}\ \bibnamefont {Wang}}, \bibinfo {author} {\bibfnamefont {S.-M.}\ \bibnamefont {Fei}},\ and\ \bibinfo {author} {\bibfnamefont {Z.-J.}\ \bibnamefont {Zheng}},\ }\bibfield  {title} {\bibinfo {title} {d locally indistinguishable maximally entangled states in $\mathbb c^d\otimes \mathbb c^d$},\ }\href@noop {} {\bibfield  {journal} {\bibinfo  {journal} {Physical Review A}\ }\textbf {\bibinfo {volume} {91}},\ \bibinfo {pages} {042318} (\bibinfo {year} {2015})}\BibitemShut {NoStop}%
\bibitem [{\citenamefont {Tian}\ \emph {et~al.}(2016)\citenamefont {Tian}, \citenamefont {Yu}, \citenamefont {Gao},\ and\ \citenamefont {Wen}}]{tian2016classification}%
  \BibitemOpen
  \bibfield  {author} {\bibinfo {author} {\bibfnamefont {G.}~\bibnamefont {Tian}}, \bibinfo {author} {\bibfnamefont {S.}~\bibnamefont {Yu}}, \bibinfo {author} {\bibfnamefont {F.}~\bibnamefont {Gao}},\ and\ \bibinfo {author} {\bibfnamefont {Q.}~\bibnamefont {Wen}},\ }\bibfield  {title} {\bibinfo {title} {Classification of locally distinguishable and indistinguishable sets of maximally entangled states},\ }\href@noop {} {\bibfield  {journal} {\bibinfo  {journal} {Physical Review A}\ }\textbf {\bibinfo {volume} {94}},\ \bibinfo {pages} {052315} (\bibinfo {year} {2016})}\BibitemShut {NoStop}%
\bibitem [{\citenamefont {Lugli}\ \emph {et~al.}(2020)\citenamefont {Lugli}, \citenamefont {Perinotti},\ and\ \citenamefont {Tosini}}]{lugli2020fermionic}%
  \BibitemOpen
  \bibfield  {author} {\bibinfo {author} {\bibfnamefont {M.}~\bibnamefont {Lugli}}, \bibinfo {author} {\bibfnamefont {P.}~\bibnamefont {Perinotti}},\ and\ \bibinfo {author} {\bibfnamefont {A.}~\bibnamefont {Tosini}},\ }\bibfield  {title} {\bibinfo {title} {Fermionic state discrimination by local operations and classical communication},\ }\href@noop {} {\bibfield  {journal} {\bibinfo  {journal} {Physical Review Letters}\ }\textbf {\bibinfo {volume} {125}},\ \bibinfo {pages} {110403} (\bibinfo {year} {2020})}\BibitemShut {NoStop}%
\bibitem [{\citenamefont {Yuan}\ \emph {et~al.}(2022)\citenamefont {Yuan}, \citenamefont {Yang},\ and\ \citenamefont {Wang}}]{yuan2022finding}%
  \BibitemOpen
  \bibfield  {author} {\bibinfo {author} {\bibfnamefont {J.-T.}\ \bibnamefont {Yuan}}, \bibinfo {author} {\bibfnamefont {Y.-H.}\ \bibnamefont {Yang}},\ and\ \bibinfo {author} {\bibfnamefont {C.-H.}\ \bibnamefont {Wang}},\ }\bibfield  {title} {\bibinfo {title} {Finding out all locally indistinguishable sets of generalized bell states},\ }\href@noop {} {\bibfield  {journal} {\bibinfo  {journal} {Quantum}\ }\textbf {\bibinfo {volume} {6}},\ \bibinfo {pages} {763} (\bibinfo {year} {2022})}\BibitemShut {NoStop}%
\bibitem [{\citenamefont {Bennett}\ \emph {et~al.}(1996)\citenamefont {Bennett}, \citenamefont {Bernstein}, \citenamefont {Popescu},\ and\ \citenamefont {Schumacher}}]{bennett1996concentrating}%
  \BibitemOpen
  \bibfield  {author} {\bibinfo {author} {\bibfnamefont {C.~H.}\ \bibnamefont {Bennett}}, \bibinfo {author} {\bibfnamefont {H.~J.}\ \bibnamefont {Bernstein}}, \bibinfo {author} {\bibfnamefont {S.}~\bibnamefont {Popescu}},\ and\ \bibinfo {author} {\bibfnamefont {B.}~\bibnamefont {Schumacher}},\ }\bibfield  {title} {\bibinfo {title} {Concentrating partial entanglement by local operations},\ }\href@noop {} {\bibfield  {journal} {\bibinfo  {journal} {Physical Review A}\ }\textbf {\bibinfo {volume} {53}},\ \bibinfo {pages} {2046} (\bibinfo {year} {1996})}\BibitemShut {NoStop}%
\bibitem [{\citenamefont {Kim}\ \emph {et~al.}(2012)\citenamefont {Kim}, \citenamefont {Lee}, \citenamefont {Kwon},\ and\ \citenamefont {Kim}}]{kim2012protecting}%
  \BibitemOpen
  \bibfield  {author} {\bibinfo {author} {\bibfnamefont {Y.-S.}\ \bibnamefont {Kim}}, \bibinfo {author} {\bibfnamefont {J.-C.}\ \bibnamefont {Lee}}, \bibinfo {author} {\bibfnamefont {O.}~\bibnamefont {Kwon}},\ and\ \bibinfo {author} {\bibfnamefont {Y.-H.}\ \bibnamefont {Kim}},\ }\bibfield  {title} {\bibinfo {title} {Protecting entanglement from decoherence using weak measurement and quantum measurement reversal},\ }\href@noop {} {\bibfield  {journal} {\bibinfo  {journal} {Nature Physics}\ }\textbf {\bibinfo {volume} {8}},\ \bibinfo {pages} {117} (\bibinfo {year} {2012})}\BibitemShut {NoStop}%
\bibitem [{\citenamefont {Kim}\ \emph {et~al.}(2023)\citenamefont {Kim}, \citenamefont {Jung}, \citenamefont {Lee},\ and\ \citenamefont {Ra}}]{doi:10.1126/sciadv.adi5261}%
  \BibitemOpen
  \bibfield  {author} {\bibinfo {author} {\bibfnamefont {H.-J.}\ \bibnamefont {Kim}}, \bibinfo {author} {\bibfnamefont {J.-H.}\ \bibnamefont {Jung}}, \bibinfo {author} {\bibfnamefont {K.-J.}\ \bibnamefont {Lee}},\ and\ \bibinfo {author} {\bibfnamefont {Y.-S.}\ \bibnamefont {Ra}},\ }\bibfield  {title} {\bibinfo {title} {Recovering quantum entanglement after its certification},\ }\href {https://doi.org/10.1126/sciadv.adi5261} {\bibfield  {journal} {\bibinfo  {journal} {Science Advances}\ }\textbf {\bibinfo {volume} {9}},\ \bibinfo {pages} {eadi5261} (\bibinfo {year} {2023})}\BibitemShut {NoStop}%
\bibitem [{\citenamefont {Fuchs}\ and\ \citenamefont {Peres}(1996)}]{fuchs1996quantum}%
  \BibitemOpen
  \bibfield  {author} {\bibinfo {author} {\bibfnamefont {C.~A.}\ \bibnamefont {Fuchs}}\ and\ \bibinfo {author} {\bibfnamefont {A.}~\bibnamefont {Peres}},\ }\bibfield  {title} {\bibinfo {title} {Quantum-state disturbance versus information gain: Uncertainty relations for quantum information},\ }\href@noop {} {\bibfield  {journal} {\bibinfo  {journal} {Physical Review A}\ }\textbf {\bibinfo {volume} {53}},\ \bibinfo {pages} {2038} (\bibinfo {year} {1996})}\BibitemShut {NoStop}%
\bibitem [{\citenamefont {Banaszek}(2001)}]{banaszek2001fidelity}%
  \BibitemOpen
  \bibfield  {author} {\bibinfo {author} {\bibfnamefont {K.}~\bibnamefont {Banaszek}},\ }\bibfield  {title} {\bibinfo {title} {Fidelity balance in quantum operations},\ }\href@noop {} {\bibfield  {journal} {\bibinfo  {journal} {Physical Review Letters}\ }\textbf {\bibinfo {volume} {86}},\ \bibinfo {pages} {1366} (\bibinfo {year} {2001})}\BibitemShut {NoStop}%
\bibitem [{\citenamefont {Buscemi}\ and\ \citenamefont {Sacchi}(2006)}]{buscemi2006information}%
  \BibitemOpen
  \bibfield  {author} {\bibinfo {author} {\bibfnamefont {F.}~\bibnamefont {Buscemi}}\ and\ \bibinfo {author} {\bibfnamefont {M.~F.}\ \bibnamefont {Sacchi}},\ }\bibfield  {title} {\bibinfo {title} {Information-disturbance trade-off in quantum-state discrimination},\ }\href@noop {} {\bibfield  {journal} {\bibinfo  {journal} {Physical Review A}\ }\textbf {\bibinfo {volume} {74}},\ \bibinfo {pages} {052320} (\bibinfo {year} {2006})}\BibitemShut {NoStop}%
\bibitem [{\citenamefont {Sacchi}(2006)}]{sacchi2006information}%
  \BibitemOpen
  \bibfield  {author} {\bibinfo {author} {\bibfnamefont {M.~F.}\ \bibnamefont {Sacchi}},\ }\bibfield  {title} {\bibinfo {title} {Information-disturbance tradeoff in estimating a maximally entangled state},\ }\href@noop {} {\bibfield  {journal} {\bibinfo  {journal} {Physical Review Letters}\ }\textbf {\bibinfo {volume} {96}},\ \bibinfo {pages} {220502} (\bibinfo {year} {2006})}\BibitemShut {NoStop}%
\bibitem [{\citenamefont {Buscemi}\ \emph {et~al.}(2008)\citenamefont {Buscemi}, \citenamefont {Hayashi},\ and\ \citenamefont {Horodecki}}]{buscemi2008global}%
  \BibitemOpen
  \bibfield  {author} {\bibinfo {author} {\bibfnamefont {F.}~\bibnamefont {Buscemi}}, \bibinfo {author} {\bibfnamefont {M.}~\bibnamefont {Hayashi}},\ and\ \bibinfo {author} {\bibfnamefont {M.}~\bibnamefont {Horodecki}},\ }\bibfield  {title} {\bibinfo {title} {Global information balance in quantum measurements},\ }\href@noop {} {\bibfield  {journal} {\bibinfo  {journal} {Physical Review Letters}\ }\textbf {\bibinfo {volume} {100}},\ \bibinfo {pages} {210504} (\bibinfo {year} {2008})}\BibitemShut {NoStop}%
\bibitem [{\citenamefont {Kretschmann}\ \emph {et~al.}(2008)\citenamefont {Kretschmann}, \citenamefont {Schlingemann},\ and\ \citenamefont {Werner}}]{kretschmann2008information}%
  \BibitemOpen
  \bibfield  {author} {\bibinfo {author} {\bibfnamefont {D.}~\bibnamefont {Kretschmann}}, \bibinfo {author} {\bibfnamefont {D.}~\bibnamefont {Schlingemann}},\ and\ \bibinfo {author} {\bibfnamefont {R.~F.}\ \bibnamefont {Werner}},\ }\bibfield  {title} {\bibinfo {title} {The information-disturbance tradeoff and the continuity of stinespring's representation},\ }\href@noop {} {\bibfield  {journal} {\bibinfo  {journal} {IEEE transactions on information theory}\ }\textbf {\bibinfo {volume} {54}},\ \bibinfo {pages} {1708} (\bibinfo {year} {2008})}\BibitemShut {NoStop}%
\bibitem [{\citenamefont {Hong}\ \emph {et~al.}(2022)\citenamefont {Hong}, \citenamefont {Kim}, \citenamefont {Cho}, \citenamefont {Kim}, \citenamefont {Lee},\ and\ \citenamefont {Lim}}]{hong2022demonstration}%
  \BibitemOpen
  \bibfield  {author} {\bibinfo {author} {\bibfnamefont {S.}~\bibnamefont {Hong}}, \bibinfo {author} {\bibfnamefont {Y.-S.}\ \bibnamefont {Kim}}, \bibinfo {author} {\bibfnamefont {Y.-W.}\ \bibnamefont {Cho}}, \bibinfo {author} {\bibfnamefont {J.}~\bibnamefont {Kim}}, \bibinfo {author} {\bibfnamefont {S.-W.}\ \bibnamefont {Lee}},\ and\ \bibinfo {author} {\bibfnamefont {H.-T.}\ \bibnamefont {Lim}},\ }\bibfield  {title} {\bibinfo {title} {Demonstration of complete information trade-off in quantum measurement},\ }\href@noop {} {\bibfield  {journal} {\bibinfo  {journal} {Physical Review Letters}\ }\textbf {\bibinfo {volume} {128}},\ \bibinfo {pages} {050401} (\bibinfo {year} {2022})}\BibitemShut {NoStop}%
\bibitem [{\citenamefont {Skrzypczyk}\ and\ \citenamefont {Linden}(2019)}]{skrzypczyk2019robustness}%
  \BibitemOpen
  \bibfield  {author} {\bibinfo {author} {\bibfnamefont {P.}~\bibnamefont {Skrzypczyk}}\ and\ \bibinfo {author} {\bibfnamefont {N.}~\bibnamefont {Linden}},\ }\bibfield  {title} {\bibinfo {title} {Robustness of measurement, discrimination games, and accessible information},\ }\href@noop {} {\bibfield  {journal} {\bibinfo  {journal} {Physical Review Letters}\ }\textbf {\bibinfo {volume} {122}},\ \bibinfo {pages} {140403} (\bibinfo {year} {2019})}\BibitemShut {NoStop}%
\bibitem [{\citenamefont {Cohen}(2007)}]{cohen2007local}%
  \BibitemOpen
  \bibfield  {author} {\bibinfo {author} {\bibfnamefont {S.~M.}\ \bibnamefont {Cohen}},\ }\bibfield  {title} {\bibinfo {title} {Local distinguishability with preservation of entanglement},\ }\href@noop {} {\bibfield  {journal} {\bibinfo  {journal} {Physical Review A}\ }\textbf {\bibinfo {volume} {75}},\ \bibinfo {pages} {052313} (\bibinfo {year} {2007})}\BibitemShut {NoStop}%
\bibitem [{\citenamefont {Cirac}\ \emph {et~al.}(2001)\citenamefont {Cirac}, \citenamefont {D{\"u}r}, \citenamefont {Kraus},\ and\ \citenamefont {Lewenstein}}]{cirac2001entangling}%
  \BibitemOpen
  \bibfield  {author} {\bibinfo {author} {\bibfnamefont {J.~I.}\ \bibnamefont {Cirac}}, \bibinfo {author} {\bibfnamefont {W.}~\bibnamefont {D{\"u}r}}, \bibinfo {author} {\bibfnamefont {B.}~\bibnamefont {Kraus}},\ and\ \bibinfo {author} {\bibfnamefont {M.}~\bibnamefont {Lewenstein}},\ }\bibfield  {title} {\bibinfo {title} {Entangling operations and their implementation using a small amount of entanglement},\ }\href@noop {} {\bibfield  {journal} {\bibinfo  {journal} {Physical Review Letters}\ }\textbf {\bibinfo {volume} {86}},\ \bibinfo {pages} {544} (\bibinfo {year} {2001})}\BibitemShut {NoStop}%
\bibitem [{\citenamefont {Bandyopadhyay}\ \emph {et~al.}(2009)\citenamefont {Bandyopadhyay}, \citenamefont {Brassard}, \citenamefont {Kimmel},\ and\ \citenamefont {Wootters}}]{bandyopadhyay2009entanglement}%
  \BibitemOpen
  \bibfield  {author} {\bibinfo {author} {\bibfnamefont {S.}~\bibnamefont {Bandyopadhyay}}, \bibinfo {author} {\bibfnamefont {G.}~\bibnamefont {Brassard}}, \bibinfo {author} {\bibfnamefont {S.}~\bibnamefont {Kimmel}},\ and\ \bibinfo {author} {\bibfnamefont {W.~K.}\ \bibnamefont {Wootters}},\ }\bibfield  {title} {\bibinfo {title} {Entanglement cost of nonlocal measurements},\ }\href@noop {} {\bibfield  {journal} {\bibinfo  {journal} {Physical Review A}\ }\textbf {\bibinfo {volume} {80}},\ \bibinfo {pages} {012313} (\bibinfo {year} {2009})}\BibitemShut {NoStop}%
\bibitem [{\citenamefont {Bandyopadhyay}\ \emph {et~al.}(2015)\citenamefont {Bandyopadhyay}, \citenamefont {Cosentino}, \citenamefont {Johnston}, \citenamefont {Russo}, \citenamefont {Watrous},\ and\ \citenamefont {Yu}}]{bandyopadhyay2015limitations}%
  \BibitemOpen
  \bibfield  {author} {\bibinfo {author} {\bibfnamefont {S.}~\bibnamefont {Bandyopadhyay}}, \bibinfo {author} {\bibfnamefont {A.}~\bibnamefont {Cosentino}}, \bibinfo {author} {\bibfnamefont {N.}~\bibnamefont {Johnston}}, \bibinfo {author} {\bibfnamefont {V.}~\bibnamefont {Russo}}, \bibinfo {author} {\bibfnamefont {J.}~\bibnamefont {Watrous}},\ and\ \bibinfo {author} {\bibfnamefont {N.}~\bibnamefont {Yu}},\ }\bibfield  {title} {\bibinfo {title} {Limitations on separable measurements by convex optimization},\ }\href@noop {} {\bibfield  {journal} {\bibinfo  {journal} {IEEE Transactions on Information Theory}\ }\textbf {\bibinfo {volume} {61}},\ \bibinfo {pages} {3593} (\bibinfo {year} {2015})}\BibitemShut {NoStop}%
\bibitem [{\citenamefont {Bandyopadhyay}\ \emph {et~al.}(2016)\citenamefont {Bandyopadhyay}, \citenamefont {Halder},\ and\ \citenamefont {Nathanson}}]{bandyopadhyay2016entanglement}%
  \BibitemOpen
  \bibfield  {author} {\bibinfo {author} {\bibfnamefont {S.}~\bibnamefont {Bandyopadhyay}}, \bibinfo {author} {\bibfnamefont {S.}~\bibnamefont {Halder}},\ and\ \bibinfo {author} {\bibfnamefont {M.}~\bibnamefont {Nathanson}},\ }\bibfield  {title} {\bibinfo {title} {Entanglement as a resource for local state discrimination in multipartite systems},\ }\href@noop {} {\bibfield  {journal} {\bibinfo  {journal} {Physical Review A}\ }\textbf {\bibinfo {volume} {94}},\ \bibinfo {pages} {022311} (\bibinfo {year} {2016})}\BibitemShut {NoStop}%
\bibitem [{\citenamefont {Bandyopadhyay}\ and\ \citenamefont {Russo}(2021)}]{bandyopadhyay2021entanglement}%
  \BibitemOpen
  \bibfield  {author} {\bibinfo {author} {\bibfnamefont {S.}~\bibnamefont {Bandyopadhyay}}\ and\ \bibinfo {author} {\bibfnamefont {V.}~\bibnamefont {Russo}},\ }\bibfield  {title} {\bibinfo {title} {Entanglement cost of discriminating noisy bell states by local operations and classical communication},\ }\href@noop {} {\bibfield  {journal} {\bibinfo  {journal} {Physical Review A}\ }\textbf {\bibinfo {volume} {104}},\ \bibinfo {pages} {032429} (\bibinfo {year} {2021})}\BibitemShut {NoStop}%
\bibitem [{\citenamefont {Yu}\ \emph {et~al.}(2014)\citenamefont {Yu}, \citenamefont {Duan},\ and\ \citenamefont {Ying}}]{yu2014distinguishability}%
  \BibitemOpen
  \bibfield  {author} {\bibinfo {author} {\bibfnamefont {N.}~\bibnamefont {Yu}}, \bibinfo {author} {\bibfnamefont {R.}~\bibnamefont {Duan}},\ and\ \bibinfo {author} {\bibfnamefont {M.}~\bibnamefont {Ying}},\ }\bibfield  {title} {\bibinfo {title} {Distinguishability of quantum states by positive operator-valued measures with positive partial transpose},\ }\href@noop {} {\bibfield  {journal} {\bibinfo  {journal} {IEEE Transactions on Information Theory}\ }\textbf {\bibinfo {volume} {60}},\ \bibinfo {pages} {2069} (\bibinfo {year} {2014})}\BibitemShut {NoStop}%
\bibitem [{\citenamefont {Chitambar}\ \emph {et~al.}(2012)\citenamefont {Chitambar}, \citenamefont {Cui},\ and\ \citenamefont {Lo}}]{chitambar2012increasing}%
  \BibitemOpen
  \bibfield  {author} {\bibinfo {author} {\bibfnamefont {E.}~\bibnamefont {Chitambar}}, \bibinfo {author} {\bibfnamefont {W.}~\bibnamefont {Cui}},\ and\ \bibinfo {author} {\bibfnamefont {H.-K.}\ \bibnamefont {Lo}},\ }\bibfield  {title} {\bibinfo {title} {Increasing entanglement monotones by separable operations},\ }\href@noop {} {\bibfield  {journal} {\bibinfo  {journal} {Physical Review Letters}\ }\textbf {\bibinfo {volume} {108}},\ \bibinfo {pages} {240504} (\bibinfo {year} {2012})}\BibitemShut {NoStop}%
\bibitem [{\citenamefont {Aniello}\ and\ \citenamefont {Lupo}(2009)}]{aniello2009relation}%
  \BibitemOpen
  \bibfield  {author} {\bibinfo {author} {\bibfnamefont {P.}~\bibnamefont {Aniello}}\ and\ \bibinfo {author} {\bibfnamefont {C.}~\bibnamefont {Lupo}},\ }\bibfield  {title} {\bibinfo {title} {On the relation between schmidt coefficients and entanglement},\ }\href@noop {} {\bibfield  {journal} {\bibinfo  {journal} {Open Systems \& Information Dynamics}\ }\textbf {\bibinfo {volume} {16}},\ \bibinfo {pages} {127} (\bibinfo {year} {2009})}\BibitemShut {NoStop}%
\bibitem [{\citenamefont {Sperling}\ and\ \citenamefont {Vogel}(2011)}]{sperling2011schmidt}%
  \BibitemOpen
  \bibfield  {author} {\bibinfo {author} {\bibfnamefont {J.}~\bibnamefont {Sperling}}\ and\ \bibinfo {author} {\bibfnamefont {W.}~\bibnamefont {Vogel}},\ }\bibfield  {title} {\bibinfo {title} {The schmidt number as a universal entanglement measure},\ }\href@noop {} {\bibfield  {journal} {\bibinfo  {journal} {Physica Scripta}\ }\textbf {\bibinfo {volume} {83}},\ \bibinfo {pages} {045002} (\bibinfo {year} {2011})}\BibitemShut {NoStop}%
\bibitem [{\citenamefont {Paige}\ \emph {et~al.}(2020)\citenamefont {Paige}, \citenamefont {Kwon}, \citenamefont {Simsek}, \citenamefont {Self}, \citenamefont {Gray},\ and\ \citenamefont {Kim}}]{paige2020quantum}%
  \BibitemOpen
  \bibfield  {author} {\bibinfo {author} {\bibfnamefont {A.~J.}\ \bibnamefont {Paige}}, \bibinfo {author} {\bibfnamefont {H.}~\bibnamefont {Kwon}}, \bibinfo {author} {\bibfnamefont {S.}~\bibnamefont {Simsek}}, \bibinfo {author} {\bibfnamefont {C.~N.}\ \bibnamefont {Self}}, \bibinfo {author} {\bibfnamefont {J.}~\bibnamefont {Gray}},\ and\ \bibinfo {author} {\bibfnamefont {M.~S.}\ \bibnamefont {Kim}},\ }\bibfield  {title} {\bibinfo {title} {Quantum delocalized interactions},\ }\href@noop {} {\bibfield  {journal} {\bibinfo  {journal} {Physical Review Letters}\ }\textbf {\bibinfo {volume} {125}},\ \bibinfo {pages} {240406} (\bibinfo {year} {2020})}\BibitemShut {NoStop}%
\bibitem [{\citenamefont {Bilash}\ \emph {et~al.}(2023)\citenamefont {Bilash}, \citenamefont {Lim}, \citenamefont {Kwon}, \citenamefont {Kim}, \citenamefont {Lim}, \citenamefont {Song},\ and\ \citenamefont {Kim}}]{Bohdan}%
  \BibitemOpen
  \bibfield  {author} {\bibinfo {author} {\bibfnamefont {B.}~\bibnamefont {Bilash}}, \bibinfo {author} {\bibfnamefont {Y.}~\bibnamefont {Lim}}, \bibinfo {author} {\bibfnamefont {H.}~\bibnamefont {Kwon}}, \bibinfo {author} {\bibfnamefont {Y.}~\bibnamefont {Kim}}, \bibinfo {author} {\bibfnamefont {H.-T.}\ \bibnamefont {Lim}}, \bibinfo {author} {\bibfnamefont {W.}~\bibnamefont {Song}},\ and\ \bibinfo {author} {\bibfnamefont {Y.-S.}\ \bibnamefont {Kim}},\ }\bibfield  {title} {\bibinfo {title} {Nondestructive discrimination of bell states between distant parties},\ }\href@noop {} {\bibfield  {journal} {\bibinfo  {journal} {arXiv preprint arXiv:2309.00869}\ } (\bibinfo {year} {2023})}\BibitemShut {NoStop}%
\bibitem [{\citenamefont {Gottesman}(1997)}]{gottesman1997stabilizer}%
  \BibitemOpen
  \bibfield  {author} {\bibinfo {author} {\bibfnamefont {D.}~\bibnamefont {Gottesman}},\ }\href@noop {} {\emph {\bibinfo {title} {Stabilizer codes and quantum error correction}}}\ (\bibinfo  {publisher} {California Institute of Technology},\ \bibinfo {year} {1997})\BibitemShut {NoStop}%
\bibitem [{\citenamefont {Shalm}\ \emph {et~al.}(2015)\citenamefont {Shalm}, \citenamefont {Meyer-Scott}, \citenamefont {Christensen}, \citenamefont {Bierhorst}, \citenamefont {Wayne}, \citenamefont {Stevens}, \citenamefont {Gerrits}, \citenamefont {Glancy}, \citenamefont {Hamel}, \citenamefont {Allman} \emph {et~al.}}]{shalm2015strong}%
  \BibitemOpen
  \bibfield  {author} {\bibinfo {author} {\bibfnamefont {L.~K.}\ \bibnamefont {Shalm}}, \bibinfo {author} {\bibfnamefont {E.}~\bibnamefont {Meyer-Scott}}, \bibinfo {author} {\bibfnamefont {B.~G.}\ \bibnamefont {Christensen}}, \bibinfo {author} {\bibfnamefont {P.}~\bibnamefont {Bierhorst}}, \bibinfo {author} {\bibfnamefont {M.~A.}\ \bibnamefont {Wayne}}, \bibinfo {author} {\bibfnamefont {M.~J.}\ \bibnamefont {Stevens}}, \bibinfo {author} {\bibfnamefont {T.}~\bibnamefont {Gerrits}}, \bibinfo {author} {\bibfnamefont {S.}~\bibnamefont {Glancy}}, \bibinfo {author} {\bibfnamefont {D.~R.}\ \bibnamefont {Hamel}}, \bibinfo {author} {\bibfnamefont {M.~S.}\ \bibnamefont {Allman}}, \emph {et~al.},\ }\bibfield  {title} {\bibinfo {title} {Strong loophole-free test of local realism},\ }\href@noop {} {\bibfield  {journal} {\bibinfo  {journal} {Physical Review Letters}\ }\textbf {\bibinfo {volume} {115}},\ \bibinfo {pages} {250402} (\bibinfo {year} {2015})}\BibitemShut {NoStop}%
\bibitem [{\citenamefont {Bowles}\ \emph {et~al.}(2018)\citenamefont {Bowles}, \citenamefont {{\v{S}}upi{\'c}}, \citenamefont {Cavalcanti},\ and\ \citenamefont {Ac{\'\i}n}}]{bowles2018device}%
  \BibitemOpen
  \bibfield  {author} {\bibinfo {author} {\bibfnamefont {J.}~\bibnamefont {Bowles}}, \bibinfo {author} {\bibfnamefont {I.}~\bibnamefont {{\v{S}}upi{\'c}}}, \bibinfo {author} {\bibfnamefont {D.}~\bibnamefont {Cavalcanti}},\ and\ \bibinfo {author} {\bibfnamefont {A.}~\bibnamefont {Ac{\'\i}n}},\ }\bibfield  {title} {\bibinfo {title} {Device-independent entanglement certification of all entangled states},\ }\href@noop {} {\bibfield  {journal} {\bibinfo  {journal} {Physical Review Letters}\ }\textbf {\bibinfo {volume} {121}},\ \bibinfo {pages} {180503} (\bibinfo {year} {2018})}\BibitemShut {NoStop}%
\bibitem [{\citenamefont {Werner}(1989)}]{werner1989quantum}%
  \BibitemOpen
  \bibfield  {author} {\bibinfo {author} {\bibfnamefont {R.~F.}\ \bibnamefont {Werner}},\ }\bibfield  {title} {\bibinfo {title} {Quantum states with einstein-podolsky-rosen correlations admitting a hidden-variable model},\ }\href {https://doi.org/10.1103/PhysRevA.40.4277} {\bibfield  {journal} {\bibinfo  {journal} {Physical Review A}\ }\textbf {\bibinfo {volume} {40}},\ \bibinfo {pages} {4277} (\bibinfo {year} {1989})}\BibitemShut {NoStop}%
\bibitem [{\citenamefont {Barrett}(2002)}]{barret2002nonsequential}%
  \BibitemOpen
  \bibfield  {author} {\bibinfo {author} {\bibfnamefont {J.}~\bibnamefont {Barrett}},\ }\bibfield  {title} {\bibinfo {title} {Nonsequential positive-operator-valued measurements on entangled mixed states do not always violate a bell inequality},\ }\href {https://doi.org/10.1103/PhysRevA.65.042302} {\bibfield  {journal} {\bibinfo  {journal} {Phys. Rev. A}\ }\textbf {\bibinfo {volume} {65}},\ \bibinfo {pages} {042302} (\bibinfo {year} {2002})}\BibitemShut {NoStop}%
\bibitem [{\citenamefont {Almeida}\ \emph {et~al.}(2007)\citenamefont {Almeida}, \citenamefont {Pironio}, \citenamefont {Barrett}, \citenamefont {T\'oth},\ and\ \citenamefont {Ac\'{\i}n}}]{almeida2007noise}%
  \BibitemOpen
  \bibfield  {author} {\bibinfo {author} {\bibfnamefont {M.~L.}\ \bibnamefont {Almeida}}, \bibinfo {author} {\bibfnamefont {S.}~\bibnamefont {Pironio}}, \bibinfo {author} {\bibfnamefont {J.}~\bibnamefont {Barrett}}, \bibinfo {author} {\bibfnamefont {G.}~\bibnamefont {T\'oth}},\ and\ \bibinfo {author} {\bibfnamefont {A.}~\bibnamefont {Ac\'{\i}n}},\ }\bibfield  {title} {\bibinfo {title} {Noise robustness of the nonlocality of entangled quantum states},\ }\href {https://doi.org/10.1103/PhysRevLett.99.040403} {\bibfield  {journal} {\bibinfo  {journal} {Phys. Rev. Lett.}\ }\textbf {\bibinfo {volume} {99}},\ \bibinfo {pages} {040403} (\bibinfo {year} {2007})}\BibitemShut {NoStop}%
\bibitem [{\citenamefont {Oszmaniec}\ \emph {et~al.}(2017)\citenamefont {Oszmaniec}, \citenamefont {Guerini}, \citenamefont {Wittek},\ and\ \citenamefont {Ac\'{\i}n}}]{oszmaniec2017simulating}%
  \BibitemOpen
  \bibfield  {author} {\bibinfo {author} {\bibfnamefont {M.}~\bibnamefont {Oszmaniec}}, \bibinfo {author} {\bibfnamefont {L.}~\bibnamefont {Guerini}}, \bibinfo {author} {\bibfnamefont {P.}~\bibnamefont {Wittek}},\ and\ \bibinfo {author} {\bibfnamefont {A.}~\bibnamefont {Ac\'{\i}n}},\ }\bibfield  {title} {\bibinfo {title} {Simulating positive-operator-valued measures with projective measurements},\ }\href {https://doi.org/10.1103/PhysRevLett.119.190501} {\bibfield  {journal} {\bibinfo  {journal} {Physical Review Letters}\ }\textbf {\bibinfo {volume} {119}},\ \bibinfo {pages} {190501} (\bibinfo {year} {2017})}\BibitemShut {NoStop}%
\bibitem [{\citenamefont {Hirsch}\ \emph {et~al.}(2017)\citenamefont {Hirsch}, \citenamefont {Quintino}, \citenamefont {V{\'{e}}rtesi}, \citenamefont {Navascu{\'{e}}s},\ and\ \citenamefont {Brunner}}]{hirsch2017betterlocalhidden}%
  \BibitemOpen
  \bibfield  {author} {\bibinfo {author} {\bibfnamefont {F.}~\bibnamefont {Hirsch}}, \bibinfo {author} {\bibfnamefont {M.~T.}\ \bibnamefont {Quintino}}, \bibinfo {author} {\bibfnamefont {T.}~\bibnamefont {V{\'{e}}rtesi}}, \bibinfo {author} {\bibfnamefont {M.}~\bibnamefont {Navascu{\'{e}}s}},\ and\ \bibinfo {author} {\bibfnamefont {N.}~\bibnamefont {Brunner}},\ }\bibfield  {title} {\bibinfo {title} {Better local hidden variable models for two-qubit {W}erner states and an upper bound on the {G}rothendieck constant {$K_G(3)$}},\ }\href {https://doi.org/10.22331/q-2017-04-25-3} {\bibfield  {journal} {\bibinfo  {journal} {{Quantum}}\ }\textbf {\bibinfo {volume} {1}},\ \bibinfo {pages} {3} (\bibinfo {year} {2017})}\BibitemShut {NoStop}%
\bibitem [{\citenamefont {Hayashi}\ \emph {et~al.}(2006)\citenamefont {Hayashi}, \citenamefont {Markham}, \citenamefont {Murao}, \citenamefont {Owari},\ and\ \citenamefont {Virmani}}]{hayashi2006bounds}%
  \BibitemOpen
  \bibfield  {author} {\bibinfo {author} {\bibfnamefont {M.}~\bibnamefont {Hayashi}}, \bibinfo {author} {\bibfnamefont {D.}~\bibnamefont {Markham}}, \bibinfo {author} {\bibfnamefont {M.}~\bibnamefont {Murao}}, \bibinfo {author} {\bibfnamefont {M.}~\bibnamefont {Owari}},\ and\ \bibinfo {author} {\bibfnamefont {S.}~\bibnamefont {Virmani}},\ }\bibfield  {title} {\bibinfo {title} {Bounds on multipartite entangled orthogonal state discrimination using local operations and classical communication},\ }\href@noop {} {\bibfield  {journal} {\bibinfo  {journal} {Physical Review Letters}\ }\textbf {\bibinfo {volume} {96}},\ \bibinfo {pages} {040501} (\bibinfo {year} {2006})}\BibitemShut {NoStop}%
\bibitem [{\citenamefont {Helwig}\ \emph {et~al.}(2012)\citenamefont {Helwig}, \citenamefont {Cui}, \citenamefont {Latorre}, \citenamefont {Riera},\ and\ \citenamefont {Lo}}]{helwig2012absolute}%
  \BibitemOpen
  \bibfield  {author} {\bibinfo {author} {\bibfnamefont {W.}~\bibnamefont {Helwig}}, \bibinfo {author} {\bibfnamefont {W.}~\bibnamefont {Cui}}, \bibinfo {author} {\bibfnamefont {J.~I.}\ \bibnamefont {Latorre}}, \bibinfo {author} {\bibfnamefont {A.}~\bibnamefont {Riera}},\ and\ \bibinfo {author} {\bibfnamefont {H.-K.}\ \bibnamefont {Lo}},\ }\bibfield  {title} {\bibinfo {title} {Absolute maximal entanglement and quantum secret sharing},\ }\href@noop {} {\bibfield  {journal} {\bibinfo  {journal} {Physical Review A}\ }\textbf {\bibinfo {volume} {86}},\ \bibinfo {pages} {052335} (\bibinfo {year} {2012})}\BibitemShut {NoStop}%
\bibitem [{\citenamefont {Helwig}\ and\ \citenamefont {Cui}(2013)}]{helwig2013absolutely}%
  \BibitemOpen
  \bibfield  {author} {\bibinfo {author} {\bibfnamefont {W.}~\bibnamefont {Helwig}}\ and\ \bibinfo {author} {\bibfnamefont {W.}~\bibnamefont {Cui}},\ }\bibfield  {title} {\bibinfo {title} {Absolutely maximally entangled states: existence and applications},\ }\href@noop {} {\bibfield  {journal} {\bibinfo  {journal} {arXiv preprint arXiv:1306.2536}\ } (\bibinfo {year} {2013})}\BibitemShut {NoStop}%
\bibitem [{\citenamefont {Goyeneche}\ \emph {et~al.}(2015)\citenamefont {Goyeneche}, \citenamefont {Alsina}, \citenamefont {Latorre}, \citenamefont {Riera},\ and\ \citenamefont {{\.Z}yczkowski}}]{goyeneche2015absolutely}%
  \BibitemOpen
  \bibfield  {author} {\bibinfo {author} {\bibfnamefont {D.}~\bibnamefont {Goyeneche}}, \bibinfo {author} {\bibfnamefont {D.}~\bibnamefont {Alsina}}, \bibinfo {author} {\bibfnamefont {J.~I.}\ \bibnamefont {Latorre}}, \bibinfo {author} {\bibfnamefont {A.}~\bibnamefont {Riera}},\ and\ \bibinfo {author} {\bibfnamefont {K.}~\bibnamefont {{\.Z}yczkowski}},\ }\bibfield  {title} {\bibinfo {title} {Absolutely maximally entangled states, combinatorial designs, and multiunitary matrices},\ }\href@noop {} {\bibfield  {journal} {\bibinfo  {journal} {Physical Review A}\ }\textbf {\bibinfo {volume} {92}},\ \bibinfo {pages} {032316} (\bibinfo {year} {2015})}\BibitemShut {NoStop}%
\bibitem [{\citenamefont {Scott}(2004)}]{scott2004multipartite}%
  \BibitemOpen
  \bibfield  {author} {\bibinfo {author} {\bibfnamefont {A.~J.}\ \bibnamefont {Scott}},\ }\bibfield  {title} {\bibinfo {title} {Multipartite entanglement, quantum-error-correcting codes, and entangling power of quantum evolutions},\ }\href@noop {} {\bibfield  {journal} {\bibinfo  {journal} {Physical Review A}\ }\textbf {\bibinfo {volume} {69}},\ \bibinfo {pages} {052330} (\bibinfo {year} {2004})}\BibitemShut {NoStop}%
\bibitem [{\citenamefont {Huber}\ \emph {et~al.}(2018)\citenamefont {Huber}, \citenamefont {Eltschka}, \citenamefont {Siewert},\ and\ \citenamefont {G{\"u}hne}}]{huber2018bounds}%
  \BibitemOpen
  \bibfield  {author} {\bibinfo {author} {\bibfnamefont {F.}~\bibnamefont {Huber}}, \bibinfo {author} {\bibfnamefont {C.}~\bibnamefont {Eltschka}}, \bibinfo {author} {\bibfnamefont {J.}~\bibnamefont {Siewert}},\ and\ \bibinfo {author} {\bibfnamefont {O.}~\bibnamefont {G{\"u}hne}},\ }\bibfield  {title} {\bibinfo {title} {Bounds on absolutely maximally entangled states from shadow inequalities, and the quantum macwilliams identity},\ }\href@noop {} {\bibfield  {journal} {\bibinfo  {journal} {Journal of Physics A: Mathematical and Theoretical}\ }\textbf {\bibinfo {volume} {51}},\ \bibinfo {pages} {175301} (\bibinfo {year} {2018})}\BibitemShut {NoStop}%
\bibitem [{\citenamefont {Raissi}\ \emph {et~al.}(2018)\citenamefont {Raissi}, \citenamefont {Gogolin}, \citenamefont {Riera},\ and\ \citenamefont {Ac{\'\i}n}}]{raissi2018optimal}%
  \BibitemOpen
  \bibfield  {author} {\bibinfo {author} {\bibfnamefont {Z.}~\bibnamefont {Raissi}}, \bibinfo {author} {\bibfnamefont {C.}~\bibnamefont {Gogolin}}, \bibinfo {author} {\bibfnamefont {A.}~\bibnamefont {Riera}},\ and\ \bibinfo {author} {\bibfnamefont {A.}~\bibnamefont {Ac{\'\i}n}},\ }\bibfield  {title} {\bibinfo {title} {Optimal quantum error correcting codes from absolutely maximally entangled states},\ }\href@noop {} {\bibfield  {journal} {\bibinfo  {journal} {Journal of Physics A: Mathematical and Theoretical}\ }\textbf {\bibinfo {volume} {51}},\ \bibinfo {pages} {075301} (\bibinfo {year} {2018})}\BibitemShut {NoStop}%
\bibitem [{\citenamefont {Raissi}\ \emph {et~al.}(2020)\citenamefont {Raissi}, \citenamefont {Teixid{\'o}}, \citenamefont {Gogolin},\ and\ \citenamefont {Ac{\'\i}n}}]{raissi2020constructions}%
  \BibitemOpen
  \bibfield  {author} {\bibinfo {author} {\bibfnamefont {Z.}~\bibnamefont {Raissi}}, \bibinfo {author} {\bibfnamefont {A.}~\bibnamefont {Teixid{\'o}}}, \bibinfo {author} {\bibfnamefont {C.}~\bibnamefont {Gogolin}},\ and\ \bibinfo {author} {\bibfnamefont {A.}~\bibnamefont {Ac{\'\i}n}},\ }\bibfield  {title} {\bibinfo {title} {Constructions of k-uniform and absolutely maximally entangled states beyond maximum distance codes},\ }\href@noop {} {\bibfield  {journal} {\bibinfo  {journal} {Physical Review Research}\ }\textbf {\bibinfo {volume} {2}},\ \bibinfo {pages} {033411} (\bibinfo {year} {2020})}\BibitemShut {NoStop}%
\bibitem [{\citenamefont {Raissi}\ \emph {et~al.}(2022)\citenamefont {Raissi}, \citenamefont {Burchardt},\ and\ \citenamefont {Barnes}}]{raissi2022general}%
  \BibitemOpen
  \bibfield  {author} {\bibinfo {author} {\bibfnamefont {Z.}~\bibnamefont {Raissi}}, \bibinfo {author} {\bibfnamefont {A.}~\bibnamefont {Burchardt}},\ and\ \bibinfo {author} {\bibfnamefont {E.}~\bibnamefont {Barnes}},\ }\bibfield  {title} {\bibinfo {title} {General stabilizer approach for constructing highly entangled graph states},\ }\href@noop {} {\bibfield  {journal} {\bibinfo  {journal} {Physical Review A}\ }\textbf {\bibinfo {volume} {106}},\ \bibinfo {pages} {062424} (\bibinfo {year} {2022})}\BibitemShut {NoStop}%
\bibitem [{\citenamefont {Page}(1993)}]{page1993average}%
  \BibitemOpen
  \bibfield  {author} {\bibinfo {author} {\bibfnamefont {D.~N.}\ \bibnamefont {Page}},\ }\bibfield  {title} {\bibinfo {title} {Average entropy of a subsystem},\ }\href@noop {} {\bibfield  {journal} {\bibinfo  {journal} {Physical Review Letters}\ }\textbf {\bibinfo {volume} {71}},\ \bibinfo {pages} {1291} (\bibinfo {year} {1993})}\BibitemShut {NoStop}%
\bibitem [{\citenamefont {{\.Z}yczkowski}\ and\ \citenamefont {Sommers}(2001)}]{zyczkowski2001induced}%
  \BibitemOpen
  \bibfield  {author} {\bibinfo {author} {\bibfnamefont {K.}~\bibnamefont {{\.Z}yczkowski}}\ and\ \bibinfo {author} {\bibfnamefont {H.-J.}\ \bibnamefont {Sommers}},\ }\bibfield  {title} {\bibinfo {title} {Induced measures in the space of mixed quantum states},\ }\href@noop {} {\bibfield  {journal} {\bibinfo  {journal} {Journal of Physics A: Mathematical and General}\ }\textbf {\bibinfo {volume} {34}},\ \bibinfo {pages} {7111} (\bibinfo {year} {2001})}\BibitemShut {NoStop}%
\bibitem [{\citenamefont {Kumagai}\ and\ \citenamefont {Hayashi}(2013)}]{kumagai2013entanglement}%
  \BibitemOpen
  \bibfield  {author} {\bibinfo {author} {\bibfnamefont {W.}~\bibnamefont {Kumagai}}\ and\ \bibinfo {author} {\bibfnamefont {M.}~\bibnamefont {Hayashi}},\ }\bibfield  {title} {\bibinfo {title} {Entanglement concentration is irreversible},\ }\href@noop {} {\bibfield  {journal} {\bibinfo  {journal} {Physical Review Letters}\ }\textbf {\bibinfo {volume} {111}},\ \bibinfo {pages} {130407} (\bibinfo {year} {2013})}\BibitemShut {NoStop}%
\bibitem [{\citenamefont {Horodecki}\ \emph {et~al.}(2005)\citenamefont {Horodecki}, \citenamefont {Oppenheim},\ and\ \citenamefont {Winter}}]{horodecki2005partial}%
  \BibitemOpen
  \bibfield  {author} {\bibinfo {author} {\bibfnamefont {M.}~\bibnamefont {Horodecki}}, \bibinfo {author} {\bibfnamefont {J.}~\bibnamefont {Oppenheim}},\ and\ \bibinfo {author} {\bibfnamefont {A.}~\bibnamefont {Winter}},\ }\bibfield  {title} {\bibinfo {title} {Partial quantum information},\ }\href@noop {} {\bibfield  {journal} {\bibinfo  {journal} {Nature}\ }\textbf {\bibinfo {volume} {436}},\ \bibinfo {pages} {673} (\bibinfo {year} {2005})}\BibitemShut {NoStop}%
\bibitem [{\citenamefont {Horodecki}\ \emph {et~al.}(2007)\citenamefont {Horodecki}, \citenamefont {Oppenheim},\ and\ \citenamefont {Winter}}]{horodecki2007quantum}%
  \BibitemOpen
  \bibfield  {author} {\bibinfo {author} {\bibfnamefont {M.}~\bibnamefont {Horodecki}}, \bibinfo {author} {\bibfnamefont {J.}~\bibnamefont {Oppenheim}},\ and\ \bibinfo {author} {\bibfnamefont {A.}~\bibnamefont {Winter}},\ }\bibfield  {title} {\bibinfo {title} {Quantum state merging and negative information},\ }\href@noop {} {\bibfield  {journal} {\bibinfo  {journal} {Communications in Mathematical Physics}\ }\textbf {\bibinfo {volume} {269}},\ \bibinfo {pages} {107} (\bibinfo {year} {2007})}\BibitemShut {NoStop}%
\bibitem [{\citenamefont {Jing}(2015)}]{jing2015unitary}%
  \BibitemOpen
  \bibfield  {author} {\bibinfo {author} {\bibfnamefont {N.}~\bibnamefont {Jing}},\ }\bibfield  {title} {\bibinfo {title} {Unitary and orthogonal equivalence of sets of matrices},\ }\href@noop {} {\bibfield  {journal} {\bibinfo  {journal} {Linear Algebra and its Applications}\ }\textbf {\bibinfo {volume} {481}},\ \bibinfo {pages} {235} (\bibinfo {year} {2015})}\BibitemShut {NoStop}%
\bibitem [{\citenamefont {Kraft}(1949)}]{kraft1949device}%
  \BibitemOpen
  \bibfield  {author} {\bibinfo {author} {\bibfnamefont {L.~G.}\ \bibnamefont {Kraft}},\ }\emph {\bibinfo {title} {A device for quantizing, grouping, and coding amplitude-modulated pulses}},\ \href@noop {} {Ph.D. thesis},\ \bibinfo  {school} {Massachusetts Institute of Technology} (\bibinfo {year} {1949})\BibitemShut {NoStop}%
\bibitem [{\citenamefont {McMillan}(1956)}]{mcmillan1956two}%
  \BibitemOpen
  \bibfield  {author} {\bibinfo {author} {\bibfnamefont {B.}~\bibnamefont {McMillan}},\ }\bibfield  {title} {\bibinfo {title} {Two inequalities implied by unique decipherability},\ }\href@noop {} {\bibfield  {journal} {\bibinfo  {journal} {IRE Transactions on Information Theory}\ }\textbf {\bibinfo {volume} {2}},\ \bibinfo {pages} {115} (\bibinfo {year} {1956})}\BibitemShut {NoStop}%
\bibitem [{\citenamefont {Tomamichel}(2022)}]{tomamichel2022information}%
  \BibitemOpen
  \bibfield  {author} {\bibinfo {author} {\bibfnamefont {M.}~\bibnamefont {Tomamichel}},\ }\href@noop {} {\bibinfo {title} {Information theory and its applications}} (\bibinfo {year} {2022})\BibitemShut {NoStop}%
\bibitem [{\citenamefont {Luo}\ \emph {et~al.}(2019)\citenamefont {Luo}, \citenamefont {Zhong}, \citenamefont {Erhard}, \citenamefont {Wang}, \citenamefont {Peng}, \citenamefont {Krenn}, \citenamefont {Jiang}, \citenamefont {Li}, \citenamefont {Liu}, \citenamefont {Lu} \emph {et~al.}}]{luo2019quantum}%
  \BibitemOpen
  \bibfield  {author} {\bibinfo {author} {\bibfnamefont {Y.-H.}\ \bibnamefont {Luo}}, \bibinfo {author} {\bibfnamefont {H.-S.}\ \bibnamefont {Zhong}}, \bibinfo {author} {\bibfnamefont {M.}~\bibnamefont {Erhard}}, \bibinfo {author} {\bibfnamefont {X.-L.}\ \bibnamefont {Wang}}, \bibinfo {author} {\bibfnamefont {L.-C.}\ \bibnamefont {Peng}}, \bibinfo {author} {\bibfnamefont {M.}~\bibnamefont {Krenn}}, \bibinfo {author} {\bibfnamefont {X.}~\bibnamefont {Jiang}}, \bibinfo {author} {\bibfnamefont {L.}~\bibnamefont {Li}}, \bibinfo {author} {\bibfnamefont {N.-L.}\ \bibnamefont {Liu}}, \bibinfo {author} {\bibfnamefont {C.-Y.}\ \bibnamefont {Lu}}, \emph {et~al.},\ }\bibfield  {title} {\bibinfo {title} {Quantum teleportation in high dimensions},\ }\href@noop {} {\bibfield  {journal} {\bibinfo  {journal} {Physical review letters}\ }\textbf {\bibinfo {volume} {123}},\ \bibinfo {pages} {070505} (\bibinfo {year} {2019})}\BibitemShut {NoStop}%
\bibitem [{\citenamefont {Smith}\ and\ \citenamefont {Leung}(2006)}]{smith2006typical}%
  \BibitemOpen
  \bibfield  {author} {\bibinfo {author} {\bibfnamefont {G.}~\bibnamefont {Smith}}\ and\ \bibinfo {author} {\bibfnamefont {D.}~\bibnamefont {Leung}},\ }\bibfield  {title} {\bibinfo {title} {Typical entanglement of stabilizer states},\ }\href@noop {} {\bibfield  {journal} {\bibinfo  {journal} {Physical Review A}\ }\textbf {\bibinfo {volume} {74}},\ \bibinfo {pages} {062314} (\bibinfo {year} {2006})}\BibitemShut {NoStop}%
\end{thebibliography}%

\end{document}